\documentclass{aa}
\usepackage{natbib}
\bibpunct{(}{)}{;}{a}{}{,}
\usepackage{graphicx}
\graphicspath{{images_de_larticle/}}

\begin{document}


\title{Local reionizations histories with merger tree of HII regions}
  
\author{Jonathan Chardin \inst{1} \thanks{\email{jc@ast.cam.ac.uk}} \and Dominique Aubert \inst{2} \and Pierre Ocvirk \inst{2}}
\institute{ Kavli Institute for Cosmology and Institute of Astronomy, Madingley Rd, Cambridge, CB3 0HA, UK \and %
            Universit\'e de Strasbourg, CNRS UMR 7550, 11 rue de l'Universit\'e,  F-67000 Strasbourg, France}


%

   \date{Received ?? ??, 2011; Accepted ?? ??, 2011}

\titlerunning{Local reionizations}

\date{Accepted / Received }

\abstract 
{} 
{
We investigate simple properties of the initial stage of the reionization process around progenitors of galaxies, such as the extent of the 
initial HII region before its fusion with the UV background and the duration of its propagation.}
{
We use a set of four reionisation simulations with different resolutions and ionizing source prescriptions. 
By using a merger tree of HII regions we made a catalog of the HII regions properties. 
As the ionized regions undergo a major merger event, we considered that they belong to the global UV background. 
By looking at the lifetime of the region and their volume until this moment we draw typical local reionization histories as 
a function of time and investigate the relation between these histories and the halo mass progenitors of the regions. 
We then use an average mass accretion history model (AMAH) to extrapolate the halo mass inside the region 
from high z to $\mathrm{z=0}$ in order to make predictions about the past reionization histories of galaxies seen today.
}
{
We found that the later an HII region appears during the reionization period, the smaller will be their related lifetime and volume before they see the global UV background. 
Quantitatively the duration and the extent of the initial growth of an HII region is strongly dependent on the mass of the inner halo and can 
be as long as $\sim 50\%$ of the reionization epoch.
We found that the most massive is a halo today, the earlier it appears and the larger are the extension and the duration of propagation of its HII region.
Quantitative predictions differ depending on the box size or the source model: 
small simulated volumes are affected by proximity effects between HII regions and halo-based source models 
predict smaller regions and slower I-front expansion than in models using star particles as ionizing sources. 
Applying this extrapolation to Milky Way-type halos leads to a maximal extent of 1.1 Mpc/h for the initial 
HII region that established itself in $\sim 150-200\pm20$ Myrs. This is consistent with prediction made using 
constrained Local Group simulation. 
Considering halos with masses comparable to those of the Local Group (MW+ M31), 
our result suggests that statistically it has not been influenced by an external front coming from a Virgo-like cluster.
}
{}    

\maketitle


\section{Introduction}
\label{intro}

The reionization is a global transition event  that saw the neutral atomic content of the Universe changed back to an ionized state.
This transition would end between $z \sim 11$ (\citealt{2009ApJS..180..330K}) and $z \sim 6$ (\citealt{2006AJ....132..117F} and \citealt{2007AJ....134.2435W}) respectively, according to  
observations of the diffusion of CMB photons on the electrons released during reionization and the absorption features in the spectra of high-redshift quasars.
It is now a challenge to properly understand this period in order to explain the impact of radiation of the first sources and their imprints on the structure formation or the temperature evolution.

However reionization is also a local process, at least in its earliest stages. Before the large ionized patches overlap
toward the end of the process, the onset and the growth of HII regions is expected to present a scatter that depends at least partly on local properties. For instance a typical 
cosmic reionization history present a sharp drop in neutral fraction between z=9 and z=6, whereas specific regions have obviously been reionized for a few hundred millions of 
years at this stage : the history of the average neutral fraction is not necessarily representative of any local reionization history. This is especially true around sources hosted by galaxy progenitors which reionized first.
The I-front propagations, the recombination rates, the merging of small local ionized patches depend on e.g. the local source and baryons distribution as well as their local evolution. 
In a modelization context where reionization may provide an answer to e.g. the missing satellite problem, it is of prime importance to know if the rise of the 
UV flux is similar to an average cosmic background or conversely strictly constrained by the local buildup of sources and I-fronts. An example is the dwarf galaxies in the Local 
Group that present well established number and radial distributions that challenge the standard LCDM framework ( see \citealt{1999ApJ...522...82K}, \citealt{1999ApJ...524L..19M}).  
It has been suggested that a local reionization process (see e.g. \citealt{2011MNRAS.417L..93O} or \citealt{2013ApJ...777...51O}), 
with UV photons produced by a central lighthouse (such as the Milky Way or M31) could provide a radiative feedback that leads to better fits to the data than an external UV background. 
It suggests that the local group experienced a local inside-out reionization, isolated from external contributions for a sufficient time . One may then ask if this assumption is reasonable 
for the constituents of the Local Group, is it a frequent or a peculiar configuration? More generally how long does it take for a galaxy to be influenced by the cosmological UV background 
and the duration of its isolation could be crucial to the initial buildup of its stellar and baryonic content.  Hereafter we name this process \textit{local reionization}, 
for this initial growth of an HII region in isolation, before it ends up connecting with the great patches that establish the UV background.

Now, a lot of effort have been made in order to properly model the phenomena thanks to the advent of cosmological radiative transfer codes (see \citealt{iliev_cosmological_2006}, 
\citealt{iliev_cosmological_2009} and \citealt{iliev_cosmological_2009-1} for a comparison between these codes).
The radiative transfer runs can be realized on dark matter fields by directly considering the dark matter halos as ionizing source sites and the dark matter as a good tracer of the gas
(see \citealt{2006MNRAS.369.1625I}, \citealt{2006MNRAS.372..679M} and \citealt{2007MNRAS.376..534I} for example).
The runs can also be post-processed on hydrodynamical simulations that previously generate a field of star particles assumed as ionizing sources 
(see \citealt{2010ApJ...724..244A} and \citealt{2012A&A...548A...9C} among others).  
Finally, recently, some authors have begun to include the radiative transfer steps directly during the hydrodynamical evolution steps of the gas 
(see \citealt{2007ApJ...667..626K}, \citealt{2010MNRAS.407.2632H}, \citealt{2011MNRAS.414.3458W} and \citealt{2011ApJ...743..169F}).
Such new methods enable to apprehend the retro-action of the radiations on the star formation and allow to achieve more and more realistic simulation of the reionization process.

%

Previous authors have focused their investigations on the evolution of the HII regions in simulation of cosmic reionization.
Studies on static HII region fields taken at different redshifts were undertaken in order to characterize the HII region sizes 
(see \citealt{2004ApJ...613....1F}, \citealt{2006MNRAS.365..115F}, \citealt{2007ApJ...654...12Z} and \citealt{2008ApJ...681..756S}) 
or shape (see \citealt{2007MNRAS.377.1043M}, \citealt{2008MNRAS.388.1501C} for example).
Recently, \cite{2011MNRAS.413.1353F} have focused on the evolution of the HII region sizes during entire simulations with the idea to characterize the impact of the method used for the HII region detection.
Recently, in \cite{2012A&A...548A...9C} we studied such a dynamical evolution in cosmological simulation by using a merger tree of HII regions which is 
comparable to previous merger tree methodology used to investigate the assembly history of dark matter halos in simulations (see e.g. \citealt{1993MNRAS.262..627L}).
Such a merger tree aims at tracking the evolution of the properties of each HII region appeared during a simulation such as their volume or their merger rate.
We put in evidence how such a methodology help us to constrain the impact of the simulations parameters, such as the ionizing source recipes or the resolution, on the global morphology of the process.  
In essence this technique focus on the `average behavior' of individual histories ($\langle x_\mathrm{ion}(t)\rangle$) of reionizations rather than the behavior of the `history 
of the average reionization' ($\langle x_\mathrm{ion}\rangle (t)$). 

In the current paper, we propose to show how the merger tree methodology enable us to go further by characterizing the \textit{local} reionization histories induced by a single or a few galaxies. 
Our main ambition is to investigate simple properties of these initial local stages of reionization such as
\renewcommand{\labelitemi}{\textbullet}
\begin{itemize}
\item the isolation duration, i.e. how long does it take for an HII region to merge with the large ionized patches that set the UV background.
\item the final volume, i.e. how large are these patches when they become part of the overlapping process.
\end{itemize}
and track their evolution during the reionization epoch (for $z>6$) as well as their dependence on the local dark matter halos mass.

For this purpose, we ran cosmological simulations with radiative post-processing to produce realistic histories of reionization down to $z\sim6$  and we produced catalogs of 
the evolution of the HII regions properties for each of the ionized patches detected during a given simulation. 
From such a catalog, we can thus study the ionized region properties, such as their volume or lifetime before they see the UV background.

In a second time we propose to go beyond and to show how such a study allows us to link the past local reionization properties with galaxies 
that we observe today. 
We use an Average Mass Accretion History (AMAH) model derived from \cite{2002ApJ...568...52W} in order to calculate the mass $\mathrm{M_0}$ 
that the dark matter halo progenitors of the HII regions would have today at $\mathrm{z=0}$.
Then, we investigate the relation between these masses and the 
lifetime or the volume of the related ionized regions before they see the UV background. 

This paper is organized as follows: In Sect. \ref{simulation}, we present the simulation properties studied in this paper. 
Then, in Sect. \ref{tracking_HII_regions} we detail the methodology used to investigate the local reionization properties in the simulations.
In Sect. \ref{results} we present our results directly extracted from the simulations before making predictions about the reionization histories 
of galaxies seen today in Sect. \ref{local_reion_z0}.
We discuss the validity of our results and try to make some prediction about the Local Group reionization history in section \ref{discussion}. 
We finally conclude and show the prospects that naturally arise from this study in Sect. \ref{prospects}.

\section{Simulations}
\label{simulation}

The simulations used in the current analysis were produced for the investigations described in \citet{2010ApJ...724..244A} and full details can be found in this article.  
The same set of simulated data has also been used in \cite{2012A&A...548A...9C} where an extensive discussion of the UV sources models is provided. 
Therefore we will limit the description of the simulations to the most important general features.

\subsection{Gas dynamics \& radiative post-processing}

\label{gas_dyn}
The dynamics of the gas and dark matter is provided by the simulation code RAMSES that handles the related physics on adaptive meshes that increase resolution where required. 
Gas dynamics is solved thanks to a 2nd order Godunov scheme combined to an HLLC Riemann solver. Collisionless dynamics is tracked from dark matter particles according to a 
gravitational potential provided by a multi-resolution multi-grid solver. Stellar particles are generated on the fly using the methodology described by \cite{2006A&A...445....1R}. 
Initial conditions were produced with the MPGRAFIC package (\citealt{ 2008ApJS..178..179P}) according to the WMAP 5 constraints on cosmological parameters (\citealt{2009ApJS..180..330K}). 
Two box sizes were considered, 50 Mpc/h and 200 Mpc/h, to assess finite volume and resolution issues, both of them with a coarse resolution of $1024^3$ with 3 additional levels of refinements.  
In addition to a larger spatial resolution, the 50 Mpc/h simulation has a more efficient source formation  and has therefore a finer description of the overlap process during 
the reionization with smaller and more numerous HII regions. However its density field is subject to finite volume effects and cannot therefore include rare density peaks 
(and associated sources or absorbants) or large voids that may be expected from a random 50 Mpc/h cube picked out from a greater volume. Furthermore, the mean free path of UV photons 
can be as large as a few tens of comoving Mpc and large HII regions of tens of comoving Mpc radii can be found even at early stages of reionization 
(see e.g. \citealt{2006MNRAS.369.1625I}). Therefore the 50 Mpc/h simulations is rather used to analyze potential resolution effect whereas the 200 Mpc/h 
has a more realistic description of the cosmic variety of sources, densities and HII regions albeit at lower resolution.

Radiative transfer is treated as a post-processing step, using the ATON code (\citealt{ 2008MNRAS.387..295A}, \citealt{2010ApJ...724..244A}). 
The methodology relies on a moment-based description of the radiative transfer equation, following the M1 approximation that provides a simple 
and local closure relation between radiative pressure and radiative energy density. The code takes advantage of GPU acceleration to solve the 
conservative equations in an explicit fashion while satisfying a very strict Courant condition set by the speed of light. The calculations 
used here consider only a single group of ionizing photons, with a typical energy of 20 eV which assumes a 50000 K black-body spectrum for 
the sources (see also \citealt{2009A&A...495..389B}). Radiative transfer has been performed at the same resolution as the hydrodynamic coarse grid 
(i.e. $1024^3$) but the current analysis is based on degraded versions ($512^3$) of the outputs of ATON. The post-processing ran on 64-256 GPUs 
configurations of the Curie-CCRT supercomputing facility.

\subsection{Ionizing source models}
\label{source}
In this work we consider two kinds of UV source models. One relying on the RAMSES self-consistent stellar particles, 
the other on the dark matter halos present in the simulations. The difference and the similarities between the models 
are discussed in extenso in \cite{2012A&A...548A...9C}. The emissivities are described below and were chosen to have a 
good convergence between the models and resolution regarding the global reionization histories: the simulations all achieve 
half reionization by $z\sim 7.2\pm 0.2$ and full reionization by $z\sim6.2$, having produced typically 2 photons per baryons. 
Because emissivities are arbitrarily modulated to provide realistic and comparable reionization histories, these models mostly 
serve to provide locations of sources consistent with the large scale distribution of matter rather than being an absolute 
model for the sources. Overall, stellar particles are too scarce because of the lack of resolution and star formation is not 
converged: their emissivities must therefore be enhanced compared to their actual mass in order to achieve a complete reionization 
by $z\sim6$.  Halos are as expected more numerous than stellar particles (by an order of magnitude typically) and produce a greater number of low luminosity sources.  
Still, reionization histories are similar to the one obtained from stellar particles and it has been showed in \cite{2012A&A...548A...9C} 
that halos  and star based reionizations produce similar HII regions percolation histories as soon as small 'bubbles' created by small 
halos have merged. 

\renewcommand{\labelitemi}{\textbullet}
\begin{itemize}
\item{Star model
\vspace{0.25cm}

As already said, we first use the star particles generated self-consistently with the RAMSES code as ionizing sources.
The star formation criterion follows the same recipe as in \cite{2006A&A...445....1R}: above a given baryon over-density ($\delta \sim 5$ in our case), 
gas transforms into constant mass stars ($1 \times 10^{6} \, \mathrm{M_{\odot}}$ and $2 \times 10^{4} \, \mathrm{M_{\odot}}$ in 200/50 Mpc/h boxes) 
with a given efficiency ($\epsilon = 0.01$).  The raw emission of a stellar particle is 90 000 UV photons per stellar baryons over its lifetime 
(taken to be equal to 20 Myrs) and is augmented by a factor 3.8/30 for the 50/200 Mpc/h simulations to produce similar reionization histories. 
Hereafter we will refer to the associated simulations with S50 and S200.

\vspace{0.25cm}}

\item{Halo model
\vspace{0.25cm}

As an alternative we consider a simple ionizing source prescription based on dark matter halos as in \cite{2006MNRAS.369.1625I}. 
Each halo is assumed as an ionizing source with a rate of photon production $\mathrm{\dot{N}_{\gamma}}$ proportional to the halo mass $\mathrm{M}$ such as:
\begin{equation}
\mathrm{\dot{N}_{\gamma}=\alpha M}
\end{equation}
\noindent 
where $\alpha$ is the constant emissivity coefficient. We have chosen values of $\alpha = 5.9 \times 10^{43}$ and $\alpha = 3.5 \times 10^{42}$ 
photons/s/$\mathrm{M_{\odot}}$ for both 200 and 50 Mpc/h boxes, respectively. Halos were detected with the parallel FOF finder of \cite{ 2011MNRAS.410.1911C} 
with a minimal mass consisting of 10 particles, corresponding to $9.8\times10^7  \mathrm{M_\odot}$ and $6.3\times10^9  \mathrm{M_\odot}$ in the 50 and 200 Mpc/h simulations. 
Halos do not have a finite lifetime and unlike the stellar UV sources, halos come with a range of mass, hence a range of luminosities.
These parameters were chosen to produce reionization histories similar to the stellar particles driven models. 
As previously, these models will be referred as H200 and H50 for the two box sizes.

\vspace{0.25cm}}
\end{itemize}

\section{Merger Trees of HII Regions}
\label{tracking_HII_regions}

The merger tree tracks the evolution of individual HII regions: it allows to study both local reionizations and also provides 
a way to quantify the evolution and geometry of the global percolation process. 
In the current work we will mostly focus on the first aspect. The HII region merger tree methodology is fully described 
in \cite{2012A&A...548A...9C} and we will focus here on novel aspects compared to this previous work.

\subsection{Merger tree of HII regions}
\label{merger}

Building the merger tree is a two stages process.

\renewcommand{\labelitemi}{\textbullet}
\begin{itemize}
\item{FOF identification
\vspace{0.25cm}

The first step aims at identifying the different HII regions in each snapshot of the simulation with a \textit{friend-of-friends} (FOF) algorithm.
Firstly, we define a ionization criterion to decide if a cell is ionized or not. In this paper we choose that a cell is ionized if its ionization fraction 
$x \ge 0.5$. We demonstrated in \cite{2012A&A...548A...9C}  that the related HII region size distribution is almost 
unchanged when we vary this threshold. 
Second, we begin the exploration of the cosmological box and each time we encounter an ionized cell we begin an HII region identification.
The basic idea is to mark with an identification number (ID) the ionized cells belonging to the HII region being explored and to mark all the cells explored (neutral or ionized) as visited.
The algorithm proceed by scanning each time the six nearest neighbors of the ionized cells encountered and to diffuse the ID from ionized near neighbors to ionized near neighbors.
We repeat this task for all ionized cells that are still unvisited until the box is totally explored and we repeat this operation for all the snapshots of the simulation.

\vspace{0.25cm}}

\item{Merger tree
\vspace{0.25cm}

The second step consists in the construction of the merger tree itself.
Firstly, we look at a snapshot at time `$\mathrm{t}$' where the cells of an identified region are located and we look at the snapshot `$\mathrm{t+1}$' 
what is the most common ID received by these cell.
We can thus link two IDs for a same HII region between two consecutive snapshot of the simulation.
Second, we repeat this task for all the HII regions between time `$\mathrm{t}$' and `$\mathrm{t+1}$'.
Finally we reproduce these operations between all consecutive snapshots to obtain the full merger tree of the simulation.   

\vspace{0.25cm}}

\end{itemize}

\subsection{Catalog of HII regions}
\label{catalog}

From the merger tree of all simulations we have derived a catalog of the HII region properties for all HII regions appeared in the simulations.
We store for each new HII region a list of properties from the moment when the region appears until the end of the simulation.
For each of these properties we fill out a list with a new occurrence for each snapshot of the simulation in its lifetime interval.   
Here we list the properties stored in our catalog for each HII region:

\renewcommand{\labelitemi}{\textbullet}
\begin{enumerate}
\item{The list of the redshifts z}
\vspace{0.25cm}
\item{The list of the IDs}
\vspace{0.25cm}
\item{The list of the volume}
\vspace{0.25cm}
\item{The list of the number of mergers with the HII region}
\vspace{0.25cm}
\item{The list of the related volume that merge with the region}
\vspace{0.005cm}
\item{The list of the number of dark matter halos enclosed inside the region}
\vspace{0.25cm}
\item{The list of the related total halo mass enclosed inside the region}
\vspace{0.25cm}
\item{The list of the mass of the most massive halo inside the region}
\vspace{0.25cm}
\end{enumerate}

In this paper we have derived such a catalog for the four simulations described in section \ref{simulation}.
In the table \ref{tab1} we summarize the number of new regions followed in all models as well as the all features of the simulations.
As we can see, models with a box size of 200 Mpc/h suffer from a lack of HII regions and statistic compared to those with a box size of 50 Mpc/h.
Moreover the halo source models tend to present more HII regions than the star particle models, as resolution is not large enough to generate a converged 
star formation history self-consistently. On the other hand, individual halo sources are dimmer than stellar sources to produce similar reionization history, 
i.e. to produce similar numbers of UV photons until z$\sim 6$.
 
\renewcommand{\arraystretch}{1.5}
\setlength{\tabcolsep}{0.5cm}
\begin{table*}
\begin{center}
\begin{tabular}{|c||c|c|c|c|}
  \hline
  Model name & S200 & H200 & S50 & H50 \\
  \hline
  Box size [Mpc/h] & 200 & 200 & 50 & 50 \\
  \hline
  Source model &  Star particles & DM halos & Star particles & DM halos \\
  \hline
 Number of new HII regions &  2748 & 19902 & 15562 & 125408 \\
  \hline
\end{tabular}
\caption{Characteristics of the simulations studied.}
\label{tab1}
\end{center}
\end{table*}

\subsection{Local reionization evolution}
\label{local_evolution}

Our main ambition is to know if there exists some tendencies in the local HII region evolutions that could depend e.g. on the class of halo mass or on the considered redshift. 
In this first study we will focus on the HII region lifetime, i.e. the amount of time between the birth of a local ionized patch and 
the moment where it percolates with large connected ionized regions. During this time, the properties of the inner reionization depends mostly on its 
inner content, hence the `local' denomination. This initial stage is therefore driven by the specificities of the investigated region : SFR, density distribution, 
clumpiness etc... Even environmental effects can be related to some extent to the inner sources properties: for instance massive halos are likely to 
have massive neighbors, because they emerge from the same  rare event in the density field distribution. After the merging of the HII region, 
the same region of space will have access to radiation produced elsewhere and be part of the so-called 'UV background'.

\begin{figure}
   \begin{center}
      \includegraphics[width=9cm,height=5cm]{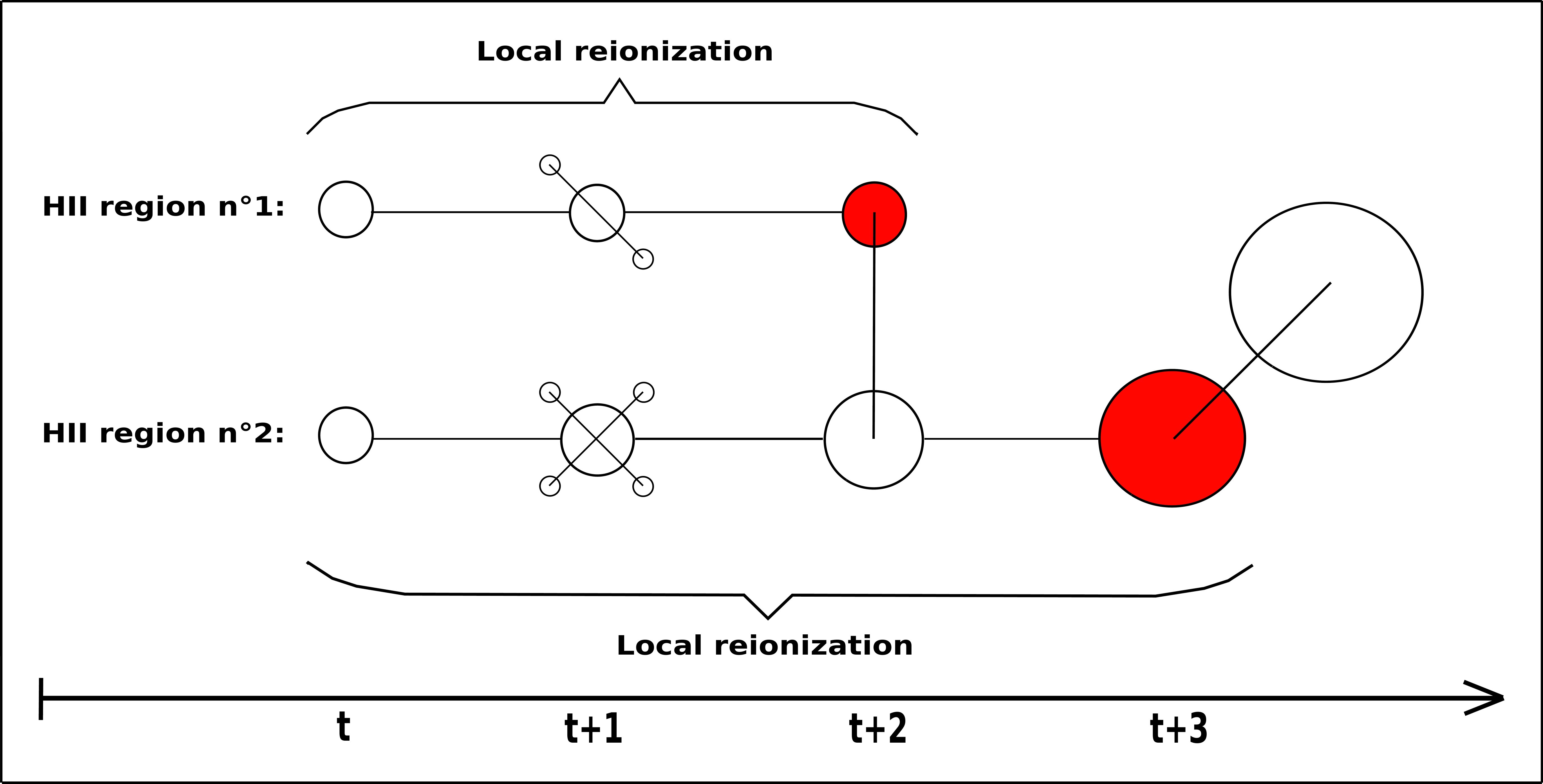} \\
  \caption{Illustration of the follow-up of the `local reionizations histories' for two HII regions.
Red items symbolize that the HII regions undergoes a major merger event with an other region larger or equal in volume.}
    \label{illustration_local_reionization}
\end{center}
\end{figure}

In this paper we have chosen that a region reaches its first major merger when the total volume of the HII regions that merge with 
the region considered is above or equal to its current volume. In figure \ref{illustration_local_reionization}, we illustrate what 
is a `local reionization history' according to our definition for two HII regions: until the region merge with one or multiple other 
regions whose the total volume is greater or equal to the current volume of the region we follow the properties of the region. 
The interval of time of this follow-up is the duration of its own `local reionization history' and we will investigate its dependence on halo mass or redshift.

\renewcommand{\arraystretch}{1.5}
\setlength{\tabcolsep}{0.5cm}
\begin{table*}[!tb]
\begin{center}
\begin{tabular}{|c|c|c|}
  \hline
  \multicolumn{3}{|c|}{200 Mpc/h} \\
\hline
   Type I & Type II & Type III  \\
  \hline
    [$\mathrm{10^{9}\le M<10^{10}}$] & [$\mathrm{10^{10}\le M<10^{11}}$] &  [$\mathrm{M \ge 10^{11}}$]\\
  \hline
 \multicolumn{3}{|c|}{50 Mpc/h} \\
\hline
   Type I & Type II & Type III  \\
  \hline
    [$\mathrm{10^{8}\le M<10^{9}}$] & [$\mathrm{10^{9}\le M<10^{10}}$] &  [$\mathrm{M \ge 10^{10}}$]\\
  \hline
\end{tabular}
\caption{Bins of dark matter halo masses at the moment of the major merger for the three HII region families.
The bins for each family are different depending on the resolution and are given for the two simulation box of 200 and 50 Mpc/h.
The mass is given in solar mass unit.}
\label{tab2}
\end{center}
\end{table*}

\section{Evolution of lifetimes and final volumes of Local reionizations}
\label{results}

First, we will present results directly extracted from the simulations.
We will only focus on the evolution of the lifetime of the HII regions before their first major merger and their volume at this moment.

\subsection{Lifetime and volume before the first major merger}
\label{ltimebmajor}

\begin{figure*}
   \begin{center}
    \begin{tabular}{cc}
      \includegraphics[width=8cm,height=7cm]{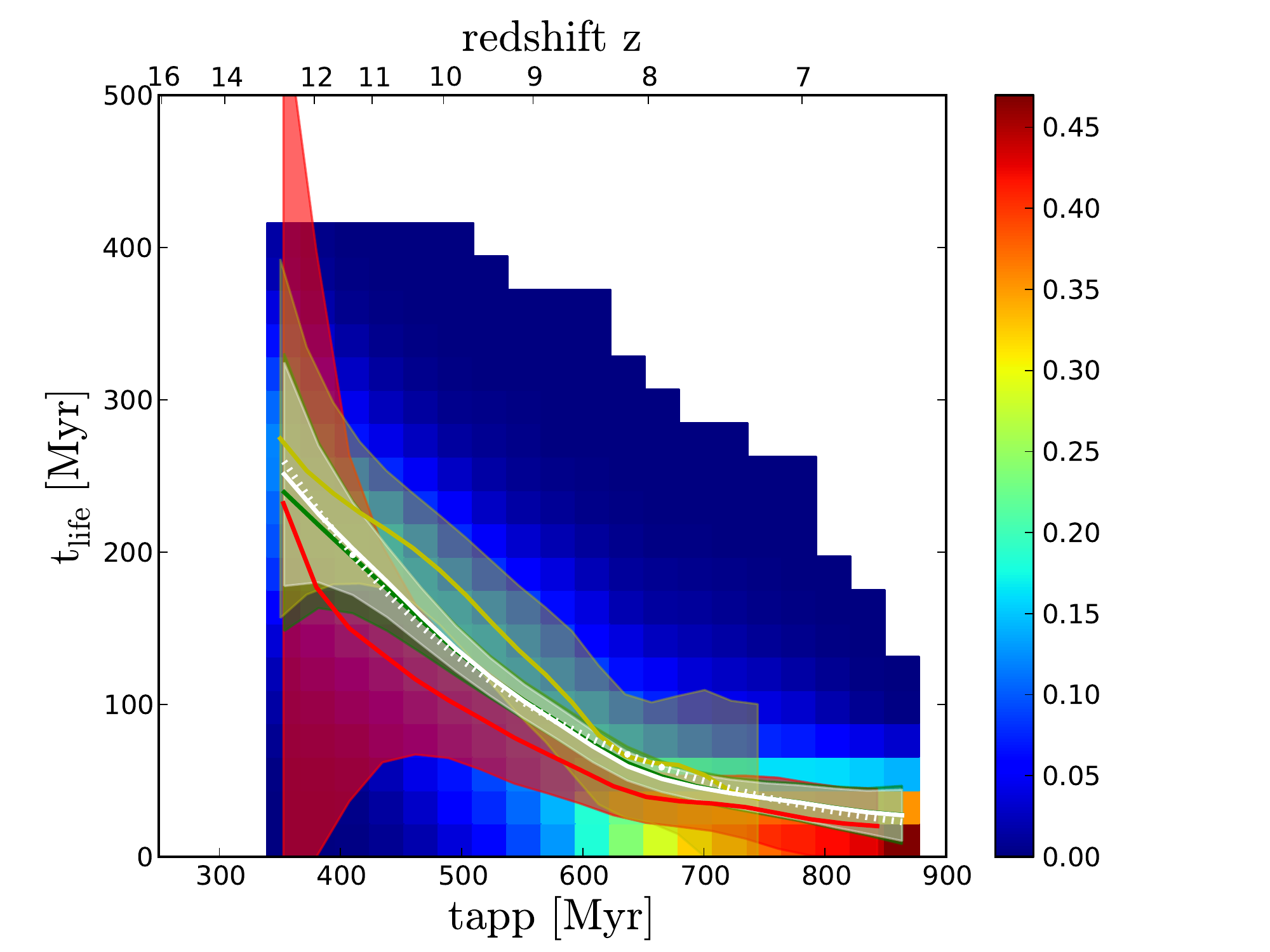} &
      \includegraphics[width=8cm,height=7cm]{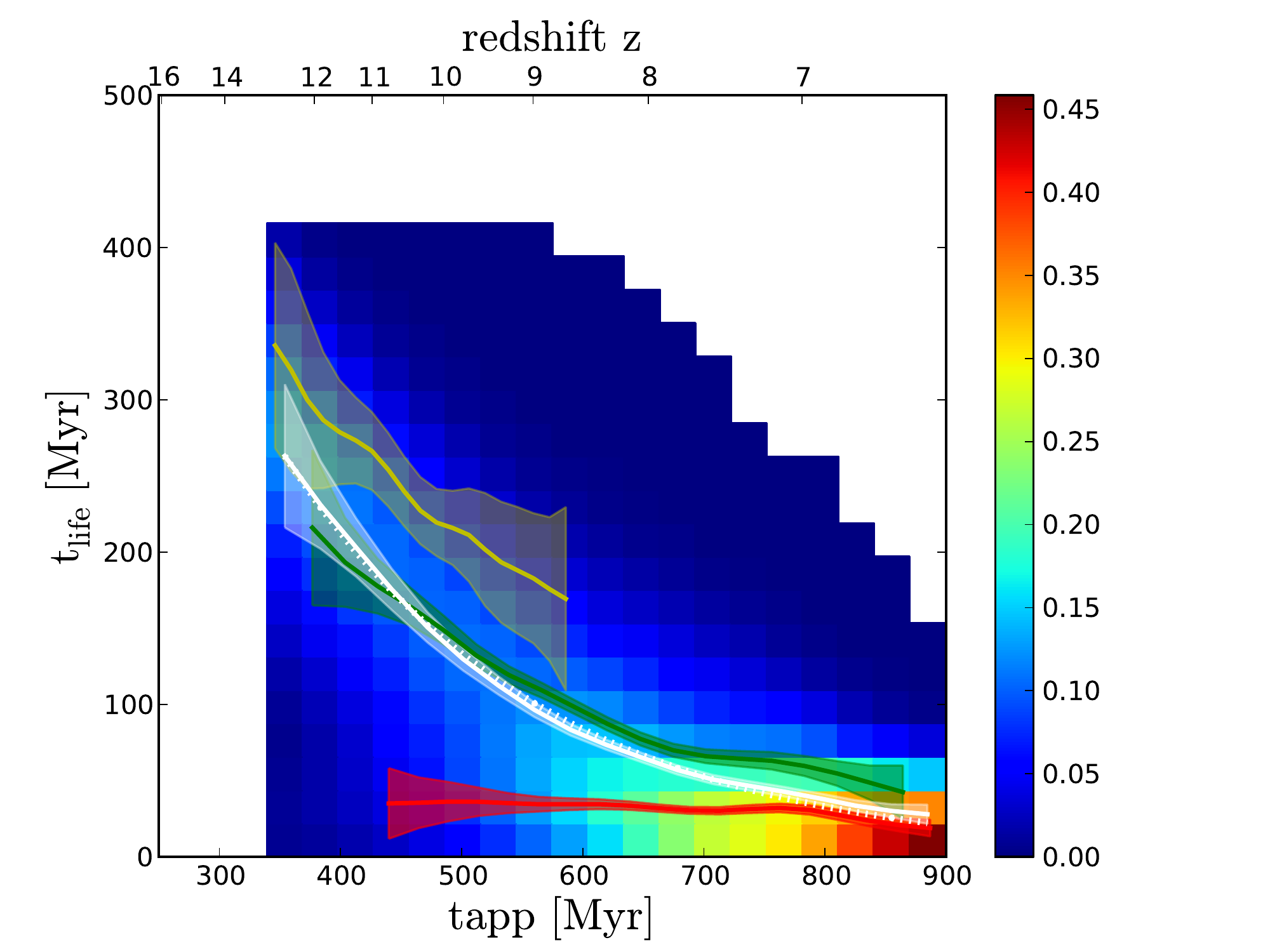} \\
      (a) S200 & (b) H200\\
      \includegraphics[width=8cm,height=7cm]{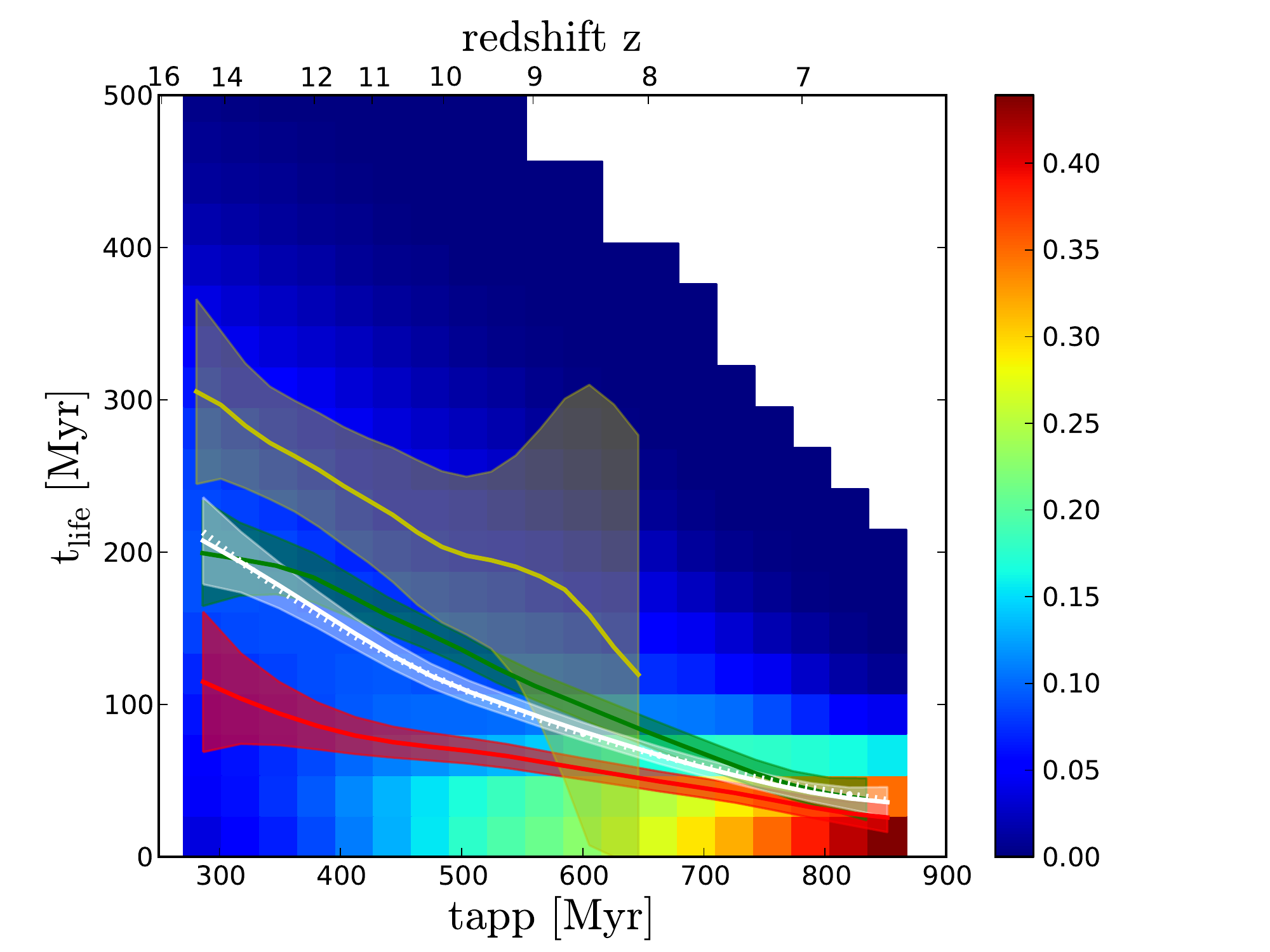} &
      \includegraphics[width=8cm,height=7cm]{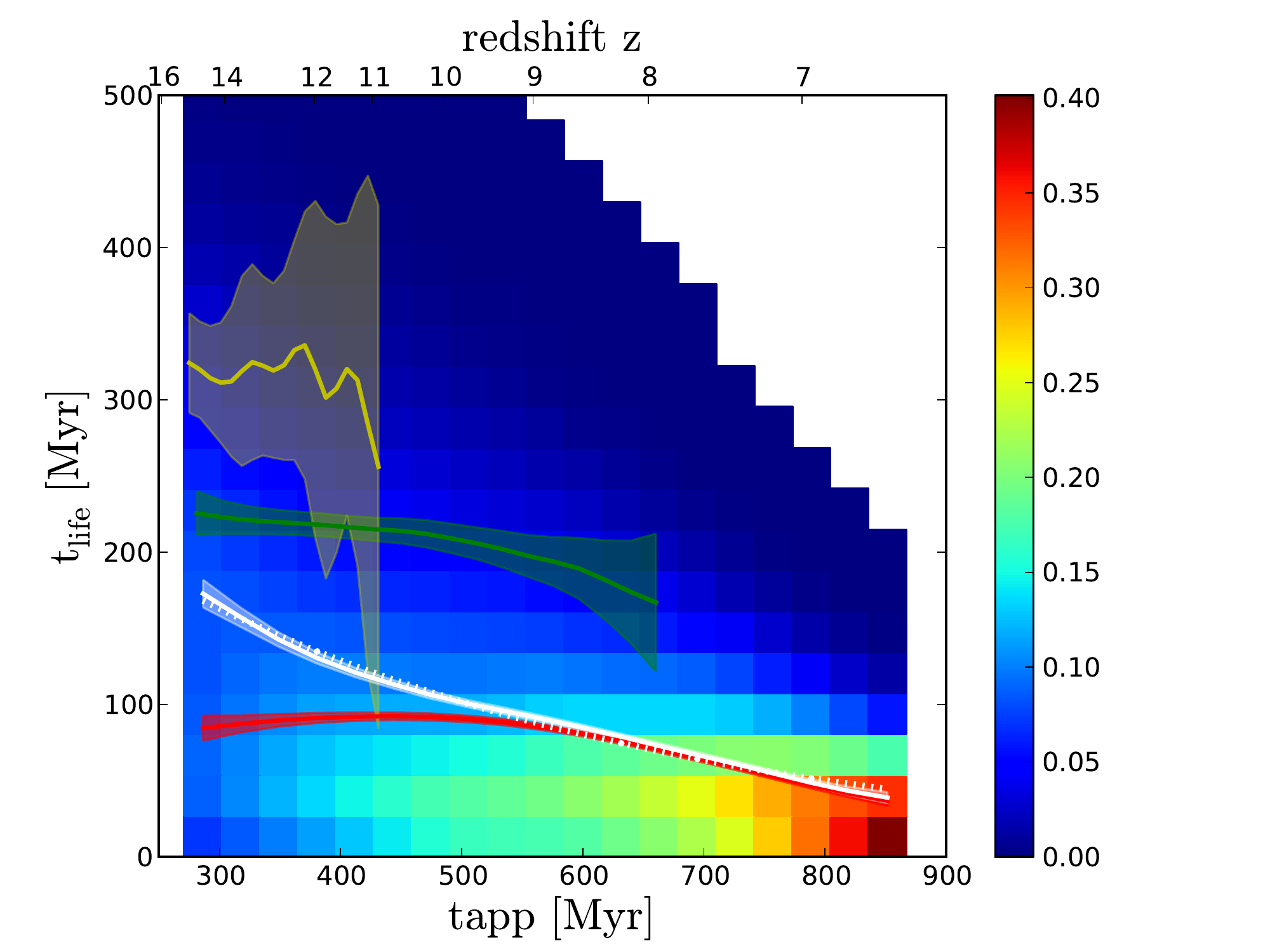} \\
       (c) S50 & (d) H50\\
\end{tabular}    
  \caption{
In background, the distribution of the lifetime of the ionized regions before their first major merger as a function of the time of apparition of the regions.
The color code is in arbitrary unit with blue values indicating a faint probability while the red tones denote a high probability.
The white curve represents the evolution of the mean value of the whole distribution and the shaded area stand for the 3$\sigma$ uncertainty on this value.
The dotted white line represents the best fit of the mean value according to the fitting formula described in appendix \ref{fit1}.
Red, green and yellow curves represent the same evolution for three classes of inner halo mass accoring to table \ref{tab2} (in solar masses for H200/S200 red: $10^9-10^{10}$, 
green : $10^{10}-10^{11}$, yellow: $>10^{11}$ and for H50/S50 red: $10^8-10^{9}$, green : $10^{9}-10^{10}$, yellow: $>10^{10}$.}
    \label{lifeteimevstime}
  \end{center}
 \end{figure*}

\begin{figure*}
   \begin{center}
    \begin{tabular}{cc}
      \includegraphics[width=8cm,height=7cm]{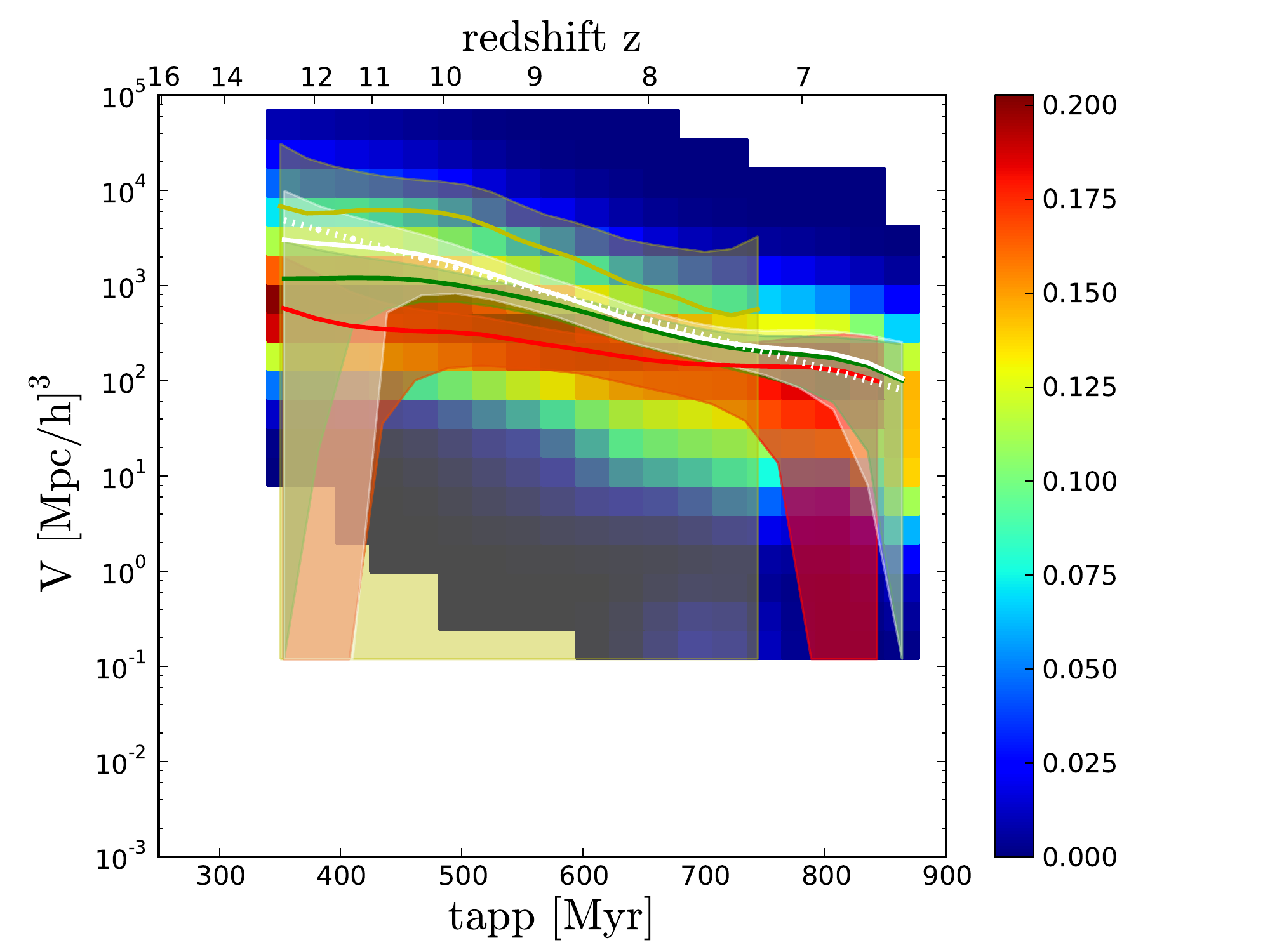} &
      \includegraphics[width=8cm,height=7cm]{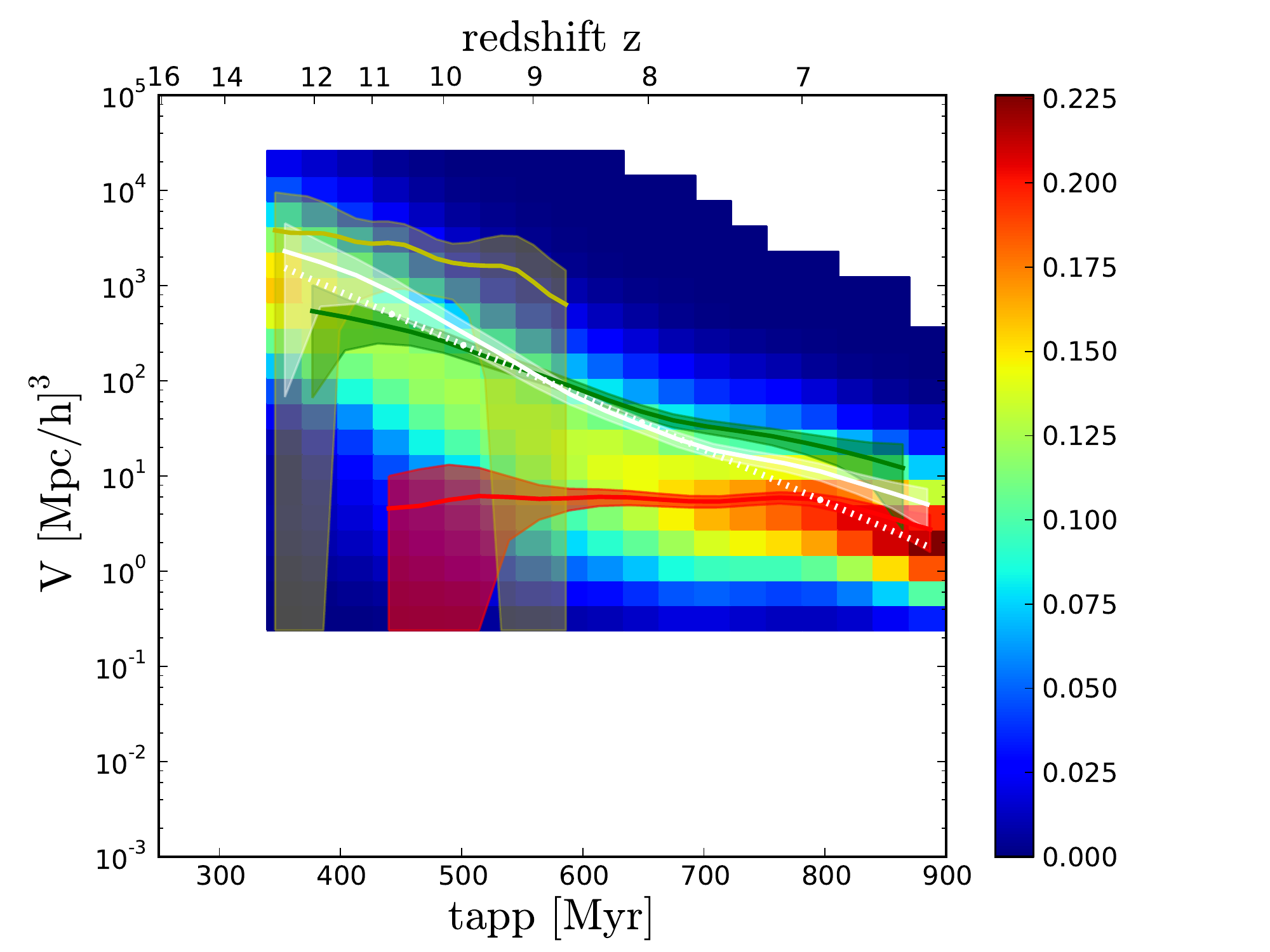} \\
      (a) S200 & (b) H200\\
      \includegraphics[width=8cm,height=7cm]{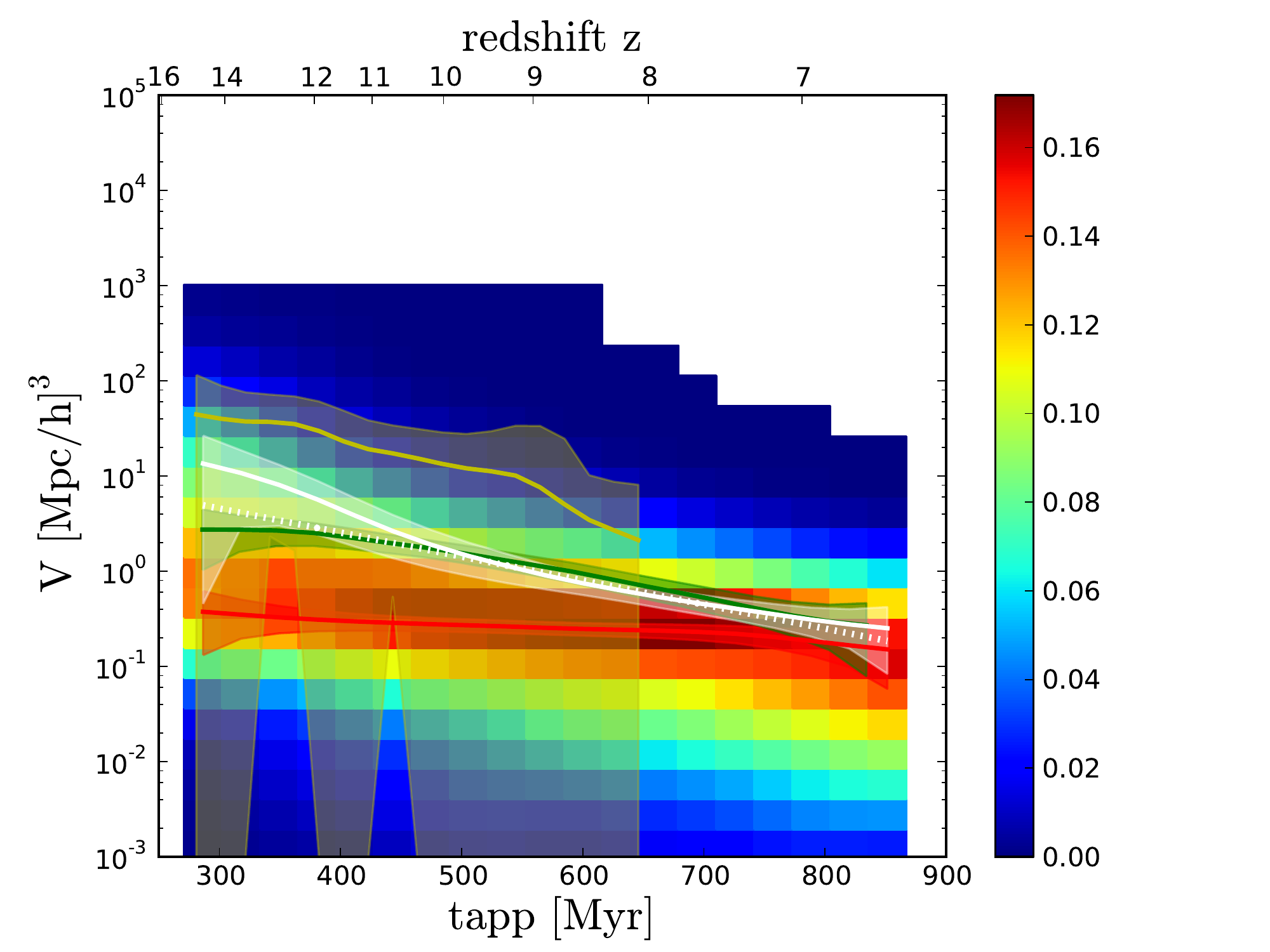} &
      \includegraphics[width=8cm,height=7cm]{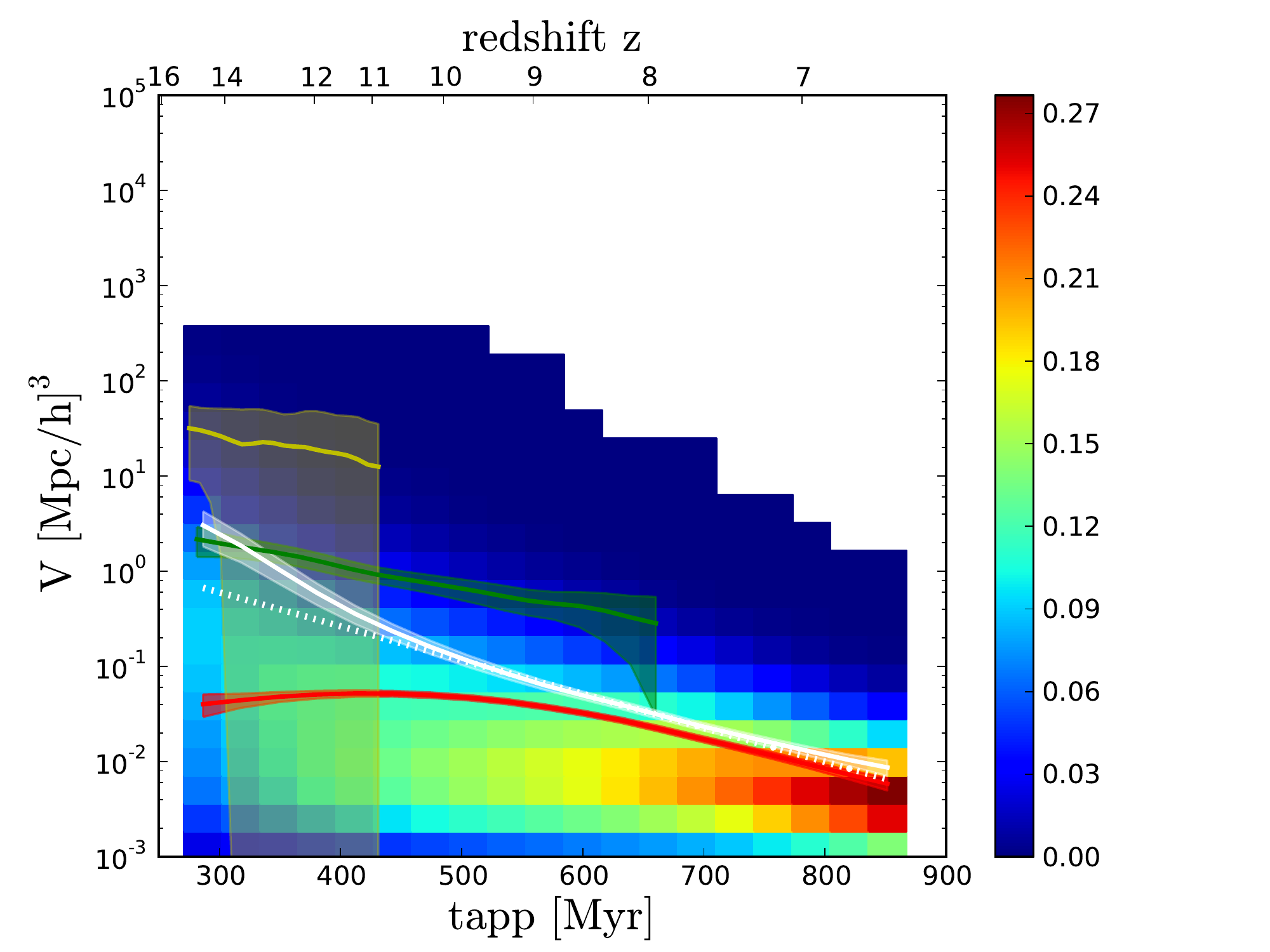} \\
      (c) S50 & (d) H50\\
\end{tabular}    
  \caption{
In background, the distribution of the volume of the ionized regions before their first major merger as a function of the time of apparition of the regions.
The color code is in arbitrary unit with blue values indicating a faint probability while the red tones denote a high probability.
The white curve represents the evolution of the mean value of the whole distribution and the shaded area stand for the 3$\sigma$ uncertainty on this value.
The dotted white line represents the best fit of the mean value according to the fitting formula described in appendix \ref{fit1}.
Red, green and yellow curves represent the same evolution for three classes of inner halo mass accoring to table \ref{tab2} (in solar masses for H200/S200 red: $10^9-10^{10}$, 
green : $10^{10}-10^{11}$, yellow: $>10^{11}$ and for H50/S50 red: $10^8-10^{9}$, green : $10^{9}-10^{10}$, yellow: $>10^{10}$.}
    \label{volume_vs_time}
  \end{center}
 \end{figure*}

Figure \ref{lifeteimevstime} presents the evolution of the distribution of the lifetime $ \mathrm{t_{life}}$ of the HII regions  
before their first major merger as a function of their cosmic time of apparition $ \mathrm{t_{app}}$. 
Figure \ref{volume_vs_time} presents the same evolution for the volume of the HII regions 
before their first major merger.
It encodes the maximal extent of the local HII region 
growth around sources. In both figures, the white curves represent  the evolution of the average of the distribution and the white shaded region stand for the $3\sigma$ uncertainty on this average.

It is reassuring to note first that the same global (and expected) evolution for the distribution is found in all simulations: 
$ \mathrm{t_{life}}$ decreases steadily with time. This drop is accompanied by a similar decrease in volume in all models.  
This is naturally explained by considering that as time passes a smaller volume remains neutral in the Universe. 
Thus, for the new regions emerging in the late phase of the reionization, the proximity effect with older regions is accentuated. 
Their life duration is thus naturally shorter than for early regions that appeared in an environment mostly neutral and their growth 
becomes also more limited.

\renewcommand{\labelitemi}{\textbullet}
\begin{itemize}
\item{Global evolution
\vspace{0.25cm}

Let us first compare source models (H50 versus S50 or H200 versus S200) : for the two box sizes, we note that the mean curves are very similar from one model of ionizing sources to another. 
At face value,  this similarity between the two ionizing source models is surprising 
considering the differences in the number of sources involved from one model to another. Indeed, we would expect to find shorter lifetime in the halo
 models because of the higher number of ionizing sources compared to the star models. This should promote proximity between the ionizing sites and 
thus encourage more rapid mergers between HII regions compared to star models if we assume that the regions grow at the same rate between the two models. 
But the models were tailored to provide similar global reionizations histories and therefore a similar number of total emitted photons is shared among a 
greater number of halos compared to star-based models. Hence we may expect slower I-fronts in H simulations and when one consider the final volumes in figure 
\ref{volume_vs_time}, one can see that H models tend to present smaller pre-merging regions than their S models counterparts: sources that are individually 
weaker in H simulations produce smaller ionized patches, with slower I-front to produce similar durations of isolation. It should also be noted that the oldest 
regions (with the smallest $ \mathrm{t_{app}}$) present similar volumes before the major merger in the two models (respectively $\sim 10^{4}$ and 
$\sim 10^{2}$ $\mathrm{Mpc^3/h}$ for the 200/50 Mpc/h box sizes). At the earliest times, regions grow in the same manner (duration and volume) 
whichever model is considered. It could imply that a good match between self-consistent stars and halos exists at this epoch, where the rarest events 
in the density field  lead to the first emitters and are less prone to numerical sub-sampling.
 
We now compare different box sizes while considering a single ionizing source model (H50 versus H200 or S50 versus S200). 
The first appearing regions have a greater lifetime at low resolution (200 Mpc/h) than at high resolution (50 Mpc/h). 
For the oldest regions  lifetimes of $\sim250-200$ Myr can be measured in S200 and S50 and $\sim300-200$ Myr for H200/H50. 
Moreover the slope of the mean curve is stronger in low resolution models than in high resolution ones.
The combination of these two facts suggests that the oldest large regions would impose a large UV background more suddenly in low 
resolution model than in high resolution ones. Indeed, when considering figure  \ref{volume_vs_time}, the final volumes of these 
primeval bubbles are larger in large simulations and are also larger in terms of fraction of the total volume. 
Large volumes produce large initial reionized patches that occupy rapidly a significant fraction of the volume and prevent 
the rise of larger and more persistent HII regions compared to smaller boxes. It could be related to the fact that S50 and 
H50 simulations are too small to have a good representation of rare events and therefore cannot produce the large, initial patches 
that drive the subsequent percolation process and that sets the duration and size of subsequent isolated HII regions. Hence,  
H50/S50 simulations present a quasi-steady decline of $\mathrm{t_{life}}$ and from one generation of new region to the next there is 
no sudden differences in their lifetime. There is a greater stationarity of the duration of local reionization histories in smaller boxes.

}

\vspace{0.25cm}
\item{Mass dependence of local reionizations
\vspace{0.25cm}

In figure \ref{lifeteimevstime} and  \ref{volume_vs_time}, we also decomposed the distribution into three sub-distributions depending on the 
mass $\mathrm{M_f}$ of the most massive halo enclosed inside the HII region at the moment of their first major merger.    
The families are defined differently regarding the box size and the summary of the family properties is given in table \ref{tab2}.  
The mean curves of the three sub-distribution at the one sigma level are shown respectively in red, green and yellow for the type I, II and III of table \ref{tab2}.

Firstly we note that in every model or box size, each class of mass occupy a dedicated location in the distributions.
We see that the greater the mass inside the region before its first major merger, the larger its lifetime $ \mathrm{t_{life}}$.
This tendency is less marked in the S200 model because of the lack of statistics compared to others models, resulting in uncertainties that overlap.
Furthermore merging HII regions with large inner mass appear among the earliest regions and cannot appear past a given cosmic time. 
If we consider volumes, HII regions with massive halos produce large isolated regions.  Overall, this behavior is expected: 
HII regions with large halos when they merge enclose objects that had sufficient time (i.e. large $ \mathrm{t_{life}}$) to accrete matter in large quantities. 
Of course large $\mathrm{t_{life}}$ are more easily obtained at early times when the ionization filling factor is still moderate. 
Conversely, regions with limited lifetime end up with small inner halos at merging and can pretty much appear at any time during 
the reionization as they are likely to produce small ionized patches.

However, it can be noted that the separate evolution of the different class of mass can be quite different than the average behaviors, especially in H models and to some extent in S50 too. 
For instance, the evolutions of the average $ \mathrm{t_{life}}$ is more abrupt than any of the evolutions in individual class of mass. 
In terms of volumes, the same discrepancy can be noted.
In these cases, the evolution of the average value is more related to the successive dominance of massive, intermediate and light inner halos. 
Because the evolutions in durations and final volumes are weaker than globally, these models show that a merging HII region of a given size contains 
a well-constrained mass and a given range of durations. It also implies that a merging ionized patch arose within a limited range of redshifts. 
}

\end{itemize}

\section{Local reionization histories as seen at z = 0}
\label{local_reion_z0}

\begin{figure}[tb]
   \begin{center}
      \includegraphics[width=9cm,height=7cm]{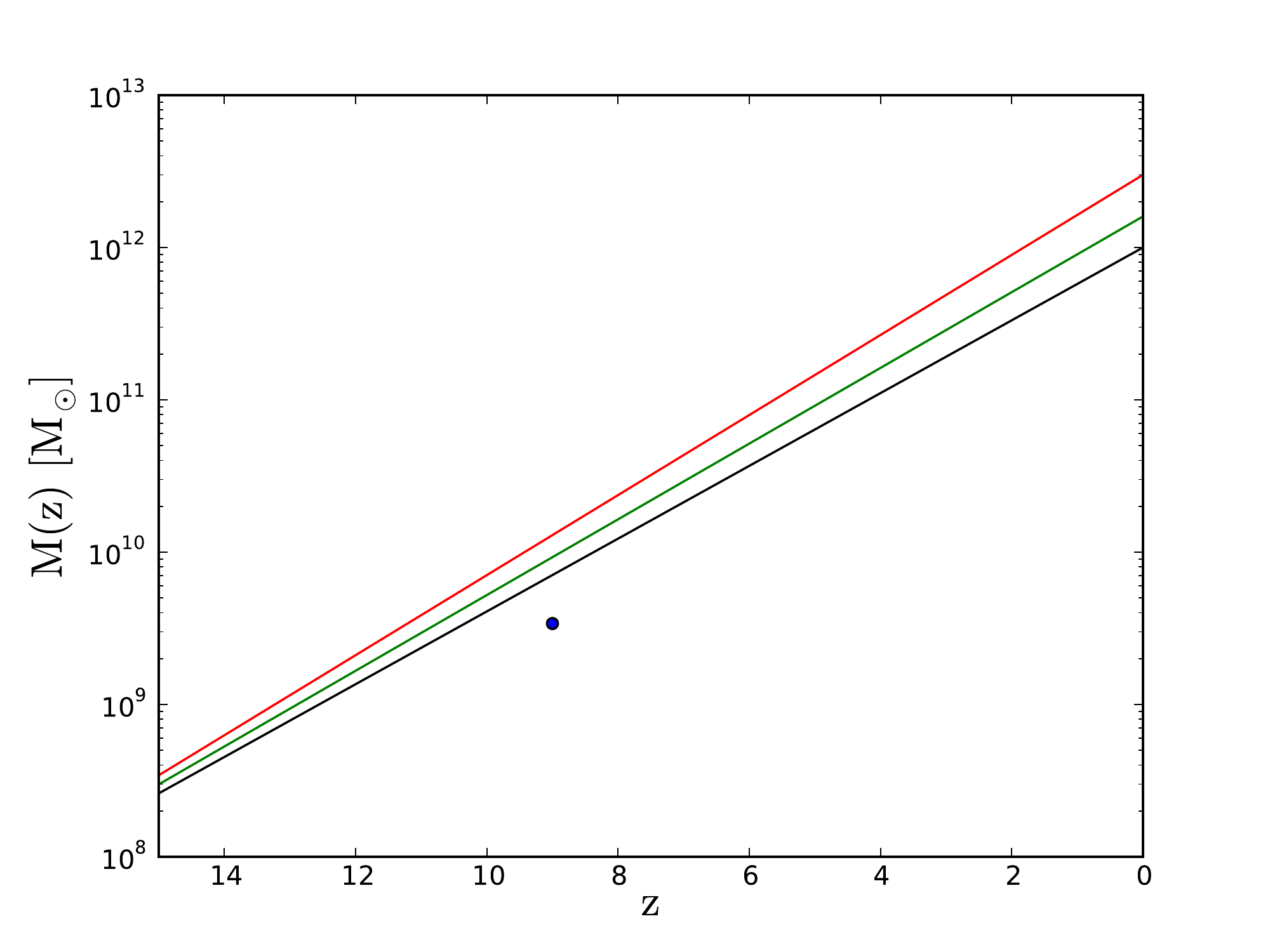} \\
  \caption{Example of dark matter halo growth calculated according to the model of \cite{2002ApJ...568...52W} (equation \ref{wechsler}) with the fits of \cite{2009MNRAS.398.1858M}.
Respectively, black, green and red curves represent the mass evolution of a MW type halo with $M_0=1\times 10^{12}  M_\odot$ as in \cite{2005MNRAS.364..433B}, 
a M31 halo with $M_0=1.6\times10^{12}  M_\odot$ as in \cite{2002ApJ...573..597K} 
and a an object that will have the mass our Local Group of galaxies today: $M_0\sim3\times 10^{12}  M_\odot$ as in \cite{2002ApJ...573..597K}.
For comparison, the blue dot stands for the mass of MW type halo at $z=9$ as found in numerical simulation by \cite{2011MNRAS.413.2093I}.
}
    \label{taux_accretion}
\end{center}
\end{figure}

Knowing the duration and extent of local reionizations at $z>6$, we aim now at extrapolating this knowledge to z=0 objects in order to have some 
insights on the past local reionization history of the galaxies observed today. For this purpose we compute the mass enclosed within HII regions 
during the epoch of reionization and extrapolate this mass from fitted halo growth relations.  Ideally, this mass would have been obtained by running 
the hydrodynamical simulations down to $z=0$ but this procedure is unfeasible at our working resolution. Details are provided in the next section, 
but in summary we used the one parameters functional form of \cite{2002ApJ...568...52W} and obtain relations such as $ \mathrm{t_{life} (M_0)}$ and $ \mathrm{V(M_0)}$. 

\subsection{Computing the mass content $\mathrm{M_0}$ inside the regions at $\mathrm{z=0}$}
\label{mzzero}
We used the 1-parameter function of \cite{2002ApJ...568...52W}:
\begin{equation}
\mathrm{ M(z)=M_0e^{-\gamma z}}
\label{wechsler}
\label{eqmcbride}
\end{equation}
that link the mass $\mathrm{M(z)}$ at redshift $\mathrm{z}$ with the mass $\mathrm{M_0}$ of the halo at redshift $\mathrm{z=0}$.
The parameter $\mathrm{\gamma}$ has the following expressions:
\begin{equation}
\mathrm{\gamma=\frac{ln(2)}{z_f}}
\end{equation}
where $\mathrm{z_f}$ is another parameter that corresponds to the redshift where the halo had the half of its actual mass $\mathrm{M(z_f)=M_0/2}$.  
We use the simulation fits of \cite{2009MNRAS.398.1858M} where the mean $\mathrm{<z_f>}$ has the following expression: 
\begin{equation}
\mathrm{<z_f>=-0.24\times \mathrm{log_{10}}\left(\frac{M_0}{10^{12}}\right)+1.26}
\end{equation}
In order to find the mass $\mathrm{M_0}$ for each HII regions we have to find the root of the following expression:
\begin{equation}
 \mathrm{M_0e^{-\gamma z}-M(z)=0}
\label{eqmcbride}
\end{equation}
with $\mathrm{M(z)}$ being the mass at a given redshift $\mathrm{z}$ that we can access in every snapshots of the simulations thanks to the catalog. 

Several choices of M(z) are possible and ideally, it would not make any difference but in practice the individual halo growth history differ from the average behavior : 
the evolution of the halo mass between the apparition and the absorption of the HII region may not follow the model used here. An obvious choice would be the mass of
the halo enclosed inside an HII region as it appears in the simulation. However an HII regions appears approximatively at the same time as the halo that hosted its 
driving source and consequently the halos have at this moment a mass close to the FOF detection threshold (10 particles) : being light in terms of particles their mass is not 
accurate and can even result in HII regions devoid of inner halos as their detection is not robust. Another choice is to consider the halo mass when the surrounding 
HII region merges with its neighbors (i.e. when $ \mathrm{t=t_{app}+t_{life}}$). At this later stages of halo buildup, masses are more important and more accurate. 
We chose to use this `final' mass $ \mathrm{M_f}$ as an input  to equation \ref{eqmcbride}, even though we also consider the less robust mass at the HII region apparition as a check. 
Finally, more than one halo can be contained inside the HII region: we decide to take the mass of the most massive object within an HII region as a realistic 
guess of the main driver of its growth.

In figure \ref{taux_accretion} we represent the evolution of the dark matter halo mass calculated with equation \ref{wechsler} for three halo with $M_0=1\times 10^{12}$ , $M_0=1.6\times 10^{12}$
and $M_0\sim3\times 10^{12}  M_\odot$ that corresponds to a MW type halo (see \citealt{2005MNRAS.364..433B}), a M31 type one and to an object that would have the mass of the Local Group at $z=0$ 
(see \citealt{2002ApJ...573..597K}). 
For comparison we show the result found for the mass of a MW-type halo at $z=9$ in numerical simulation (see \citealt{2011MNRAS.413.2093I}).
We see that the model of \cite{2002ApJ...568...52W} leads to a difference of a factor of $\sim2$ with the results of the simulation.
We even obtain the good order of magnitude and we stress that the following results are obviously model dependant and would need simulation results at $z=0$ to be more precise.

In the following we attempt to present trend about the reionization histories of galaxies observed today taking into account that our quantitative results are prone to variations.  
From the relations detailed above, and for each HII region with $\mathrm{t_{life}}$ and V  as it merges, 
we compute the $ \mathrm{t_{life}(M_0)}$ and $ \mathrm{V(M_0)}$ at z=0.

\subsection{Lifetime and volume as a function of $\mathrm{M(z=0)=M_0}$}
\label{ltimebmajor2}

\begin{figure*}
   \begin{center}
    \begin{tabular}{cc}
      \includegraphics[width=8cm,height=7cm]{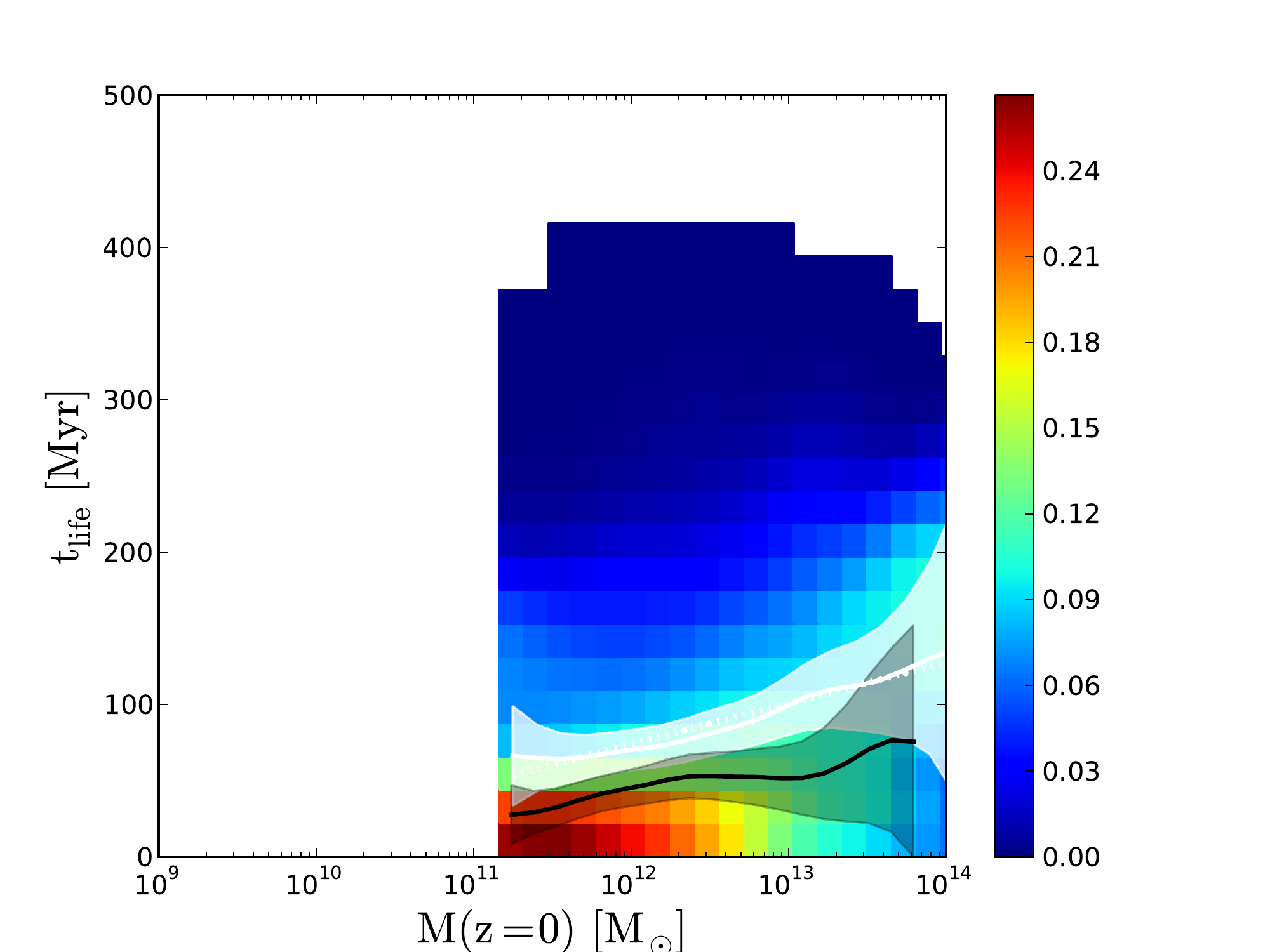} &
      \includegraphics[width=8cm,height=7cm]{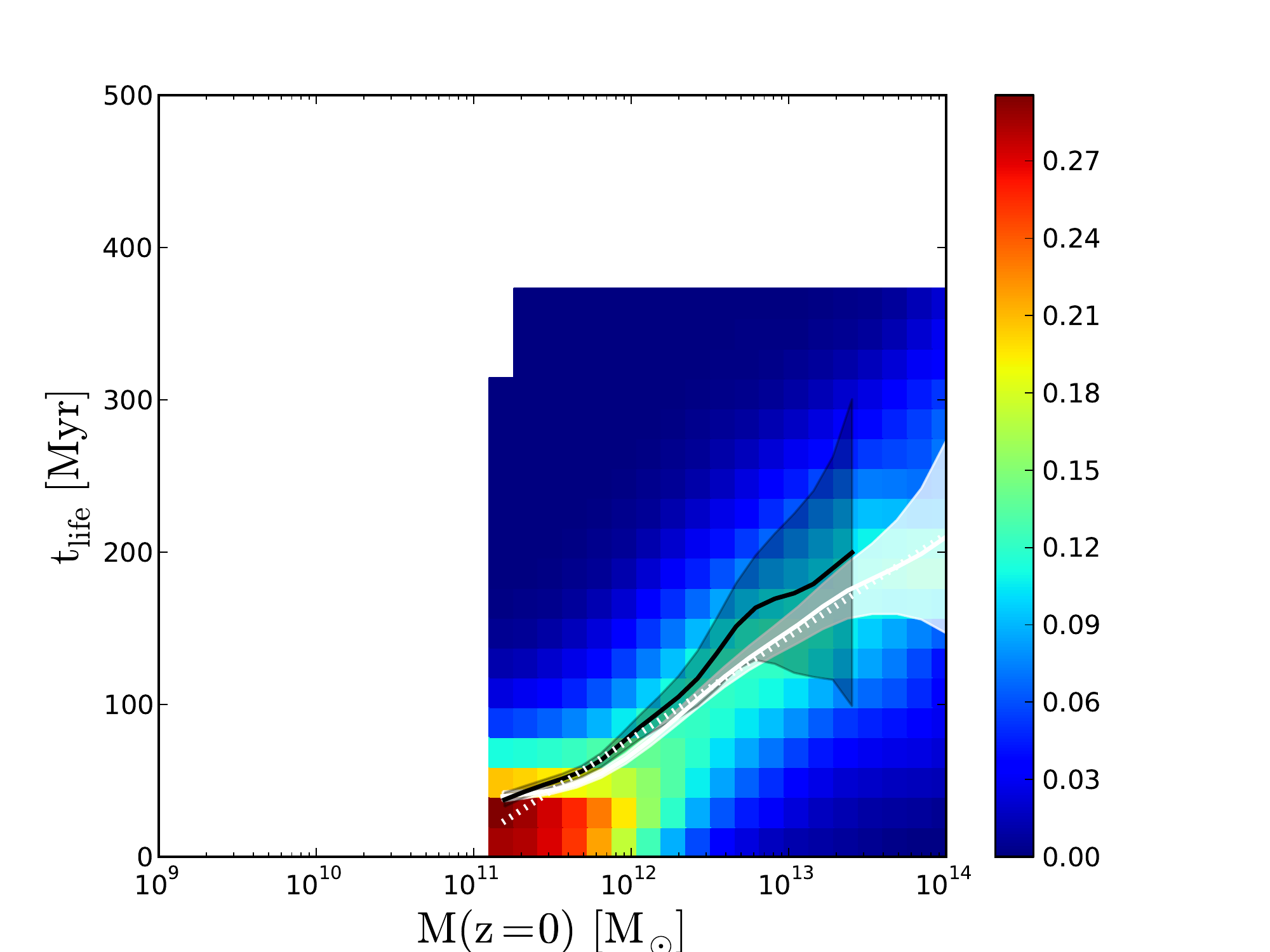} \\
      (a) S200 & (b) H200\\
      \includegraphics[width=8cm,height=7cm]{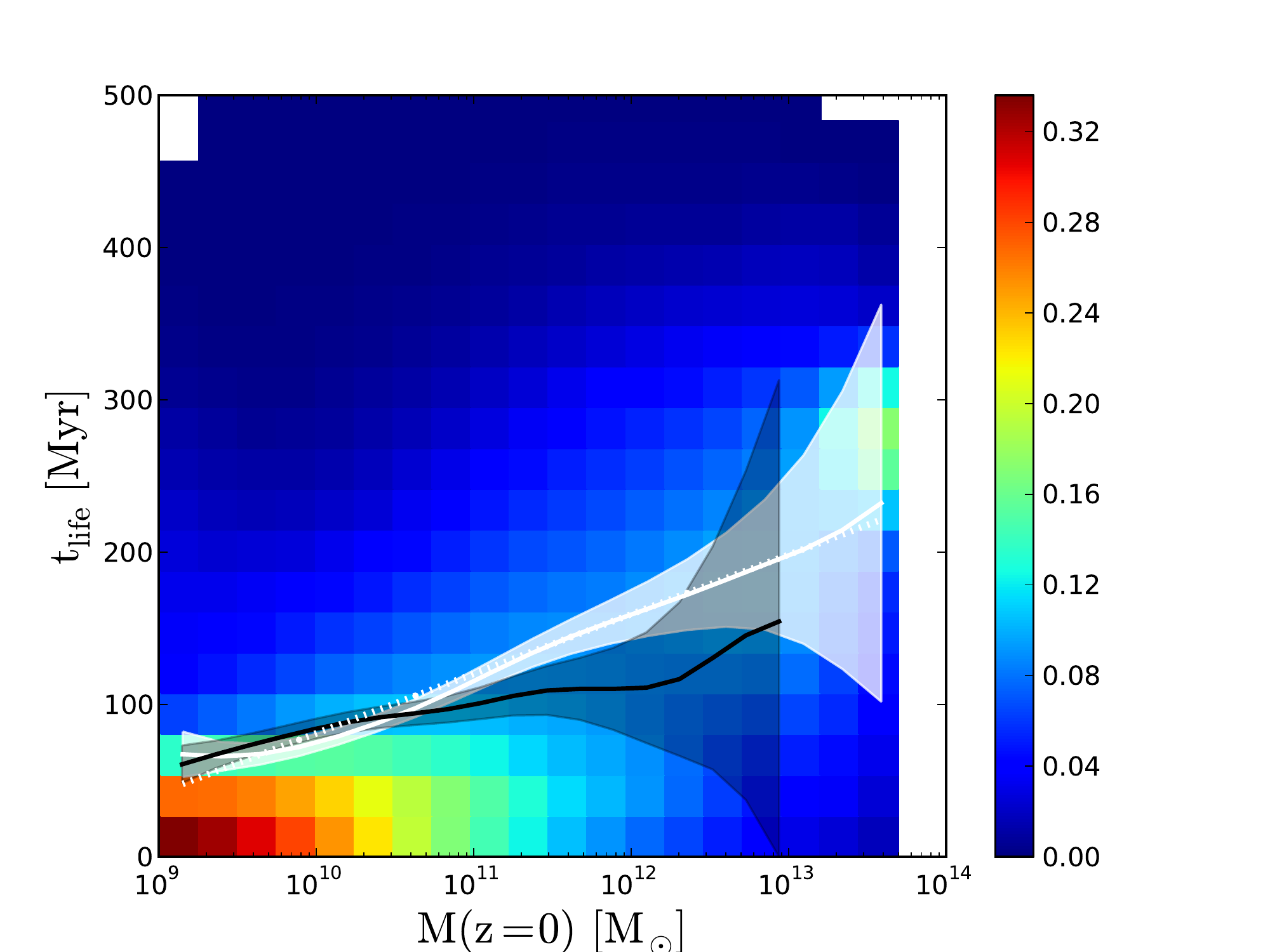} &
      \includegraphics[width=8cm,height=7cm]{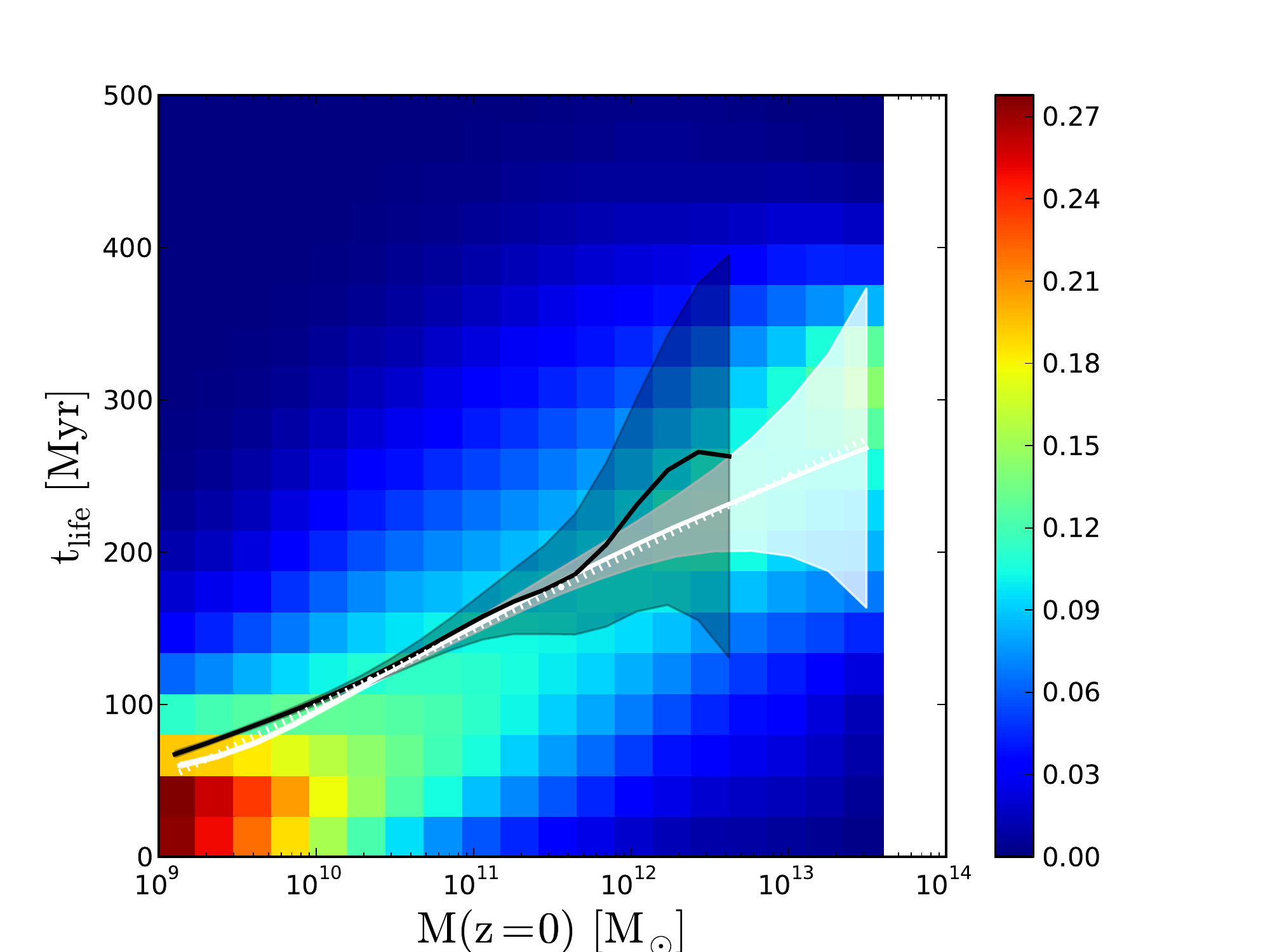} \\
       (c) S50 & (d) H50\\
\end{tabular}    
  \caption{ In background, the distribution of the duration of initial HII region growth $t_\mathrm{life}$ for a halo of current $M_0$ mass, extrapolated from 
its value when the HII region experiences a major merger. 
The color code is in arbitrary unit with blue values indicating a faint probability while the red tones denote a high probability.
The white curve represents the evolution of the mean value of the whole distribution and the shaded area stand for the 3$\sigma$ uncertainty on this value.
The dotted white line represents the best fit of the mean value according to the fitting formula described in appendix \ref{fit2}.
In black the same average relation extrapolating the halo mass from the moment the HII region appears.}
    \label{lifeteimevsmzero}
  \end{center}
 \end{figure*}

\begin{figure*}
   \begin{center}
    \begin{tabular}{cc}
      \includegraphics[width=8cm,height=7cm]{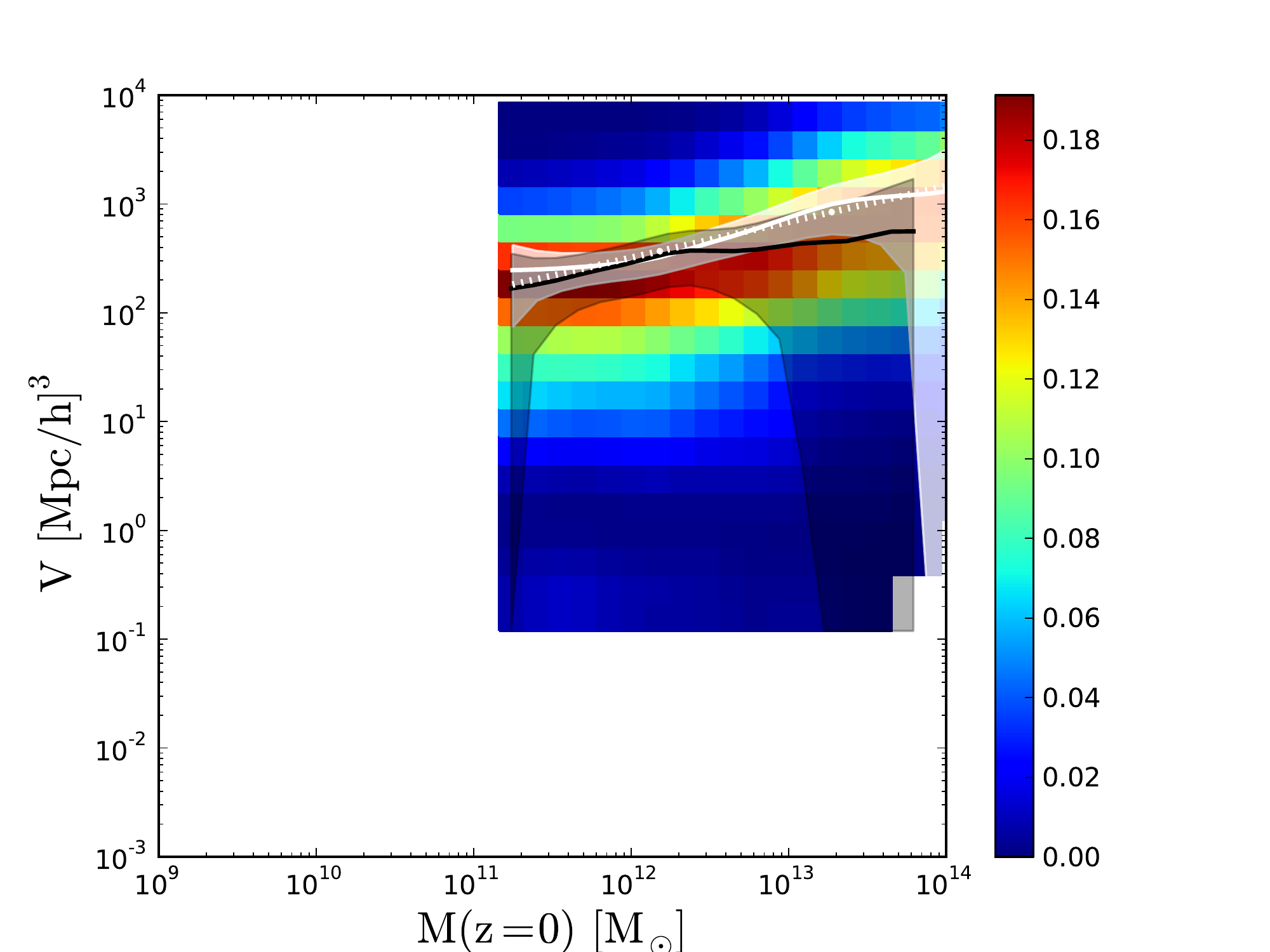} &
      \includegraphics[width=8cm,height=7cm]{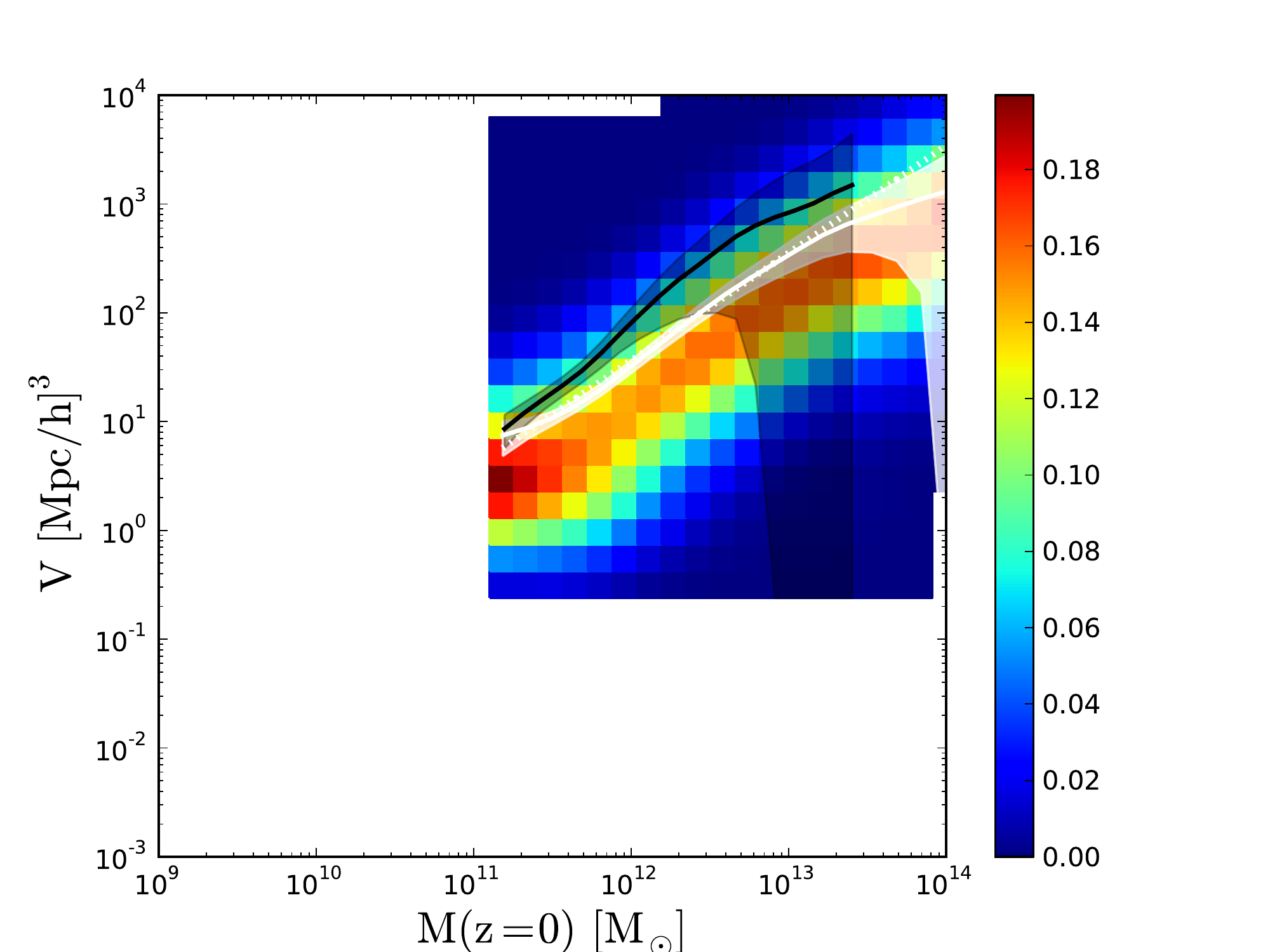} \\
      (a) S200 & (b) H200\\
      \includegraphics[width=8cm,height=7cm]{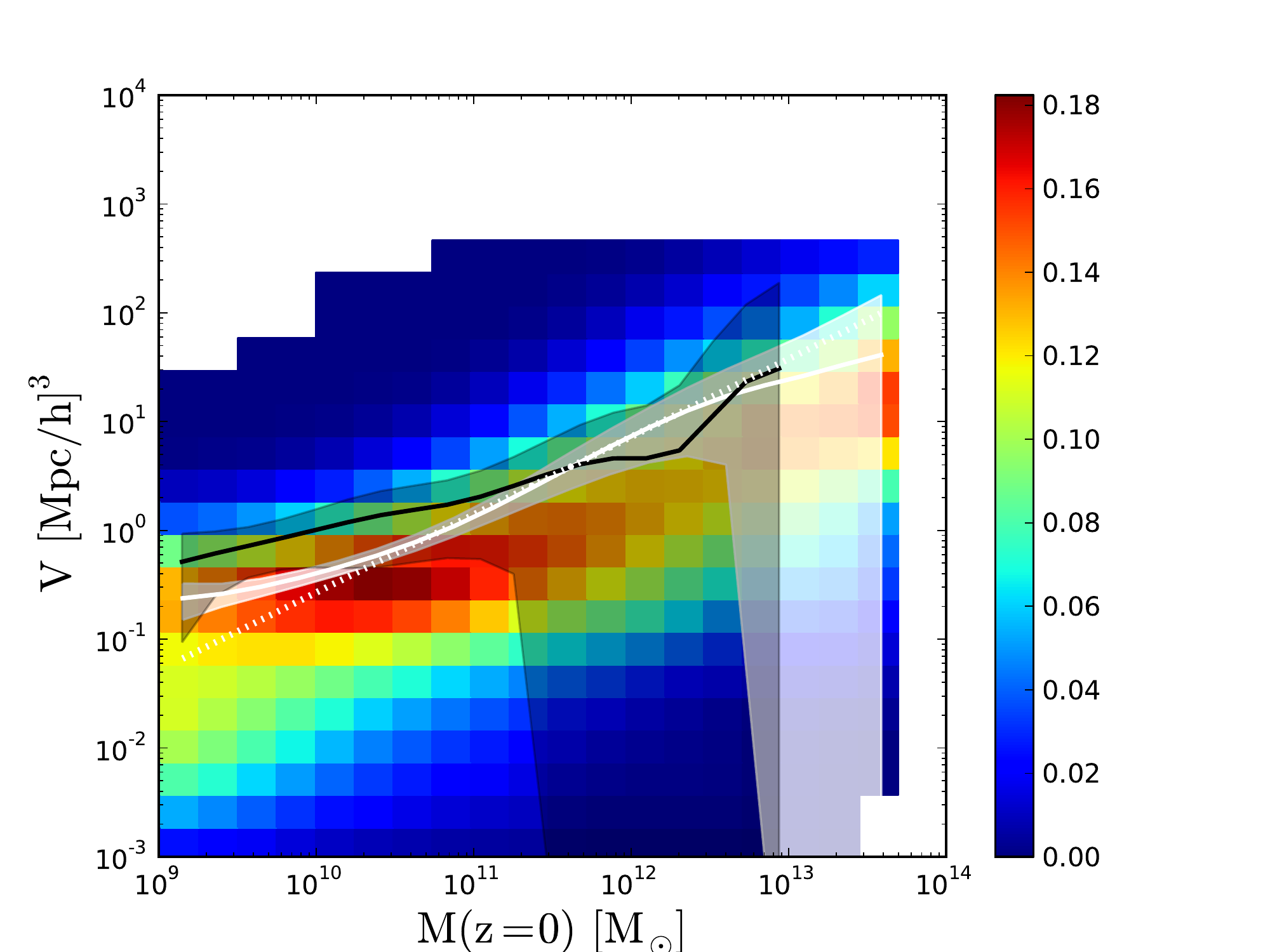} &
      \includegraphics[width=8cm,height=7cm]{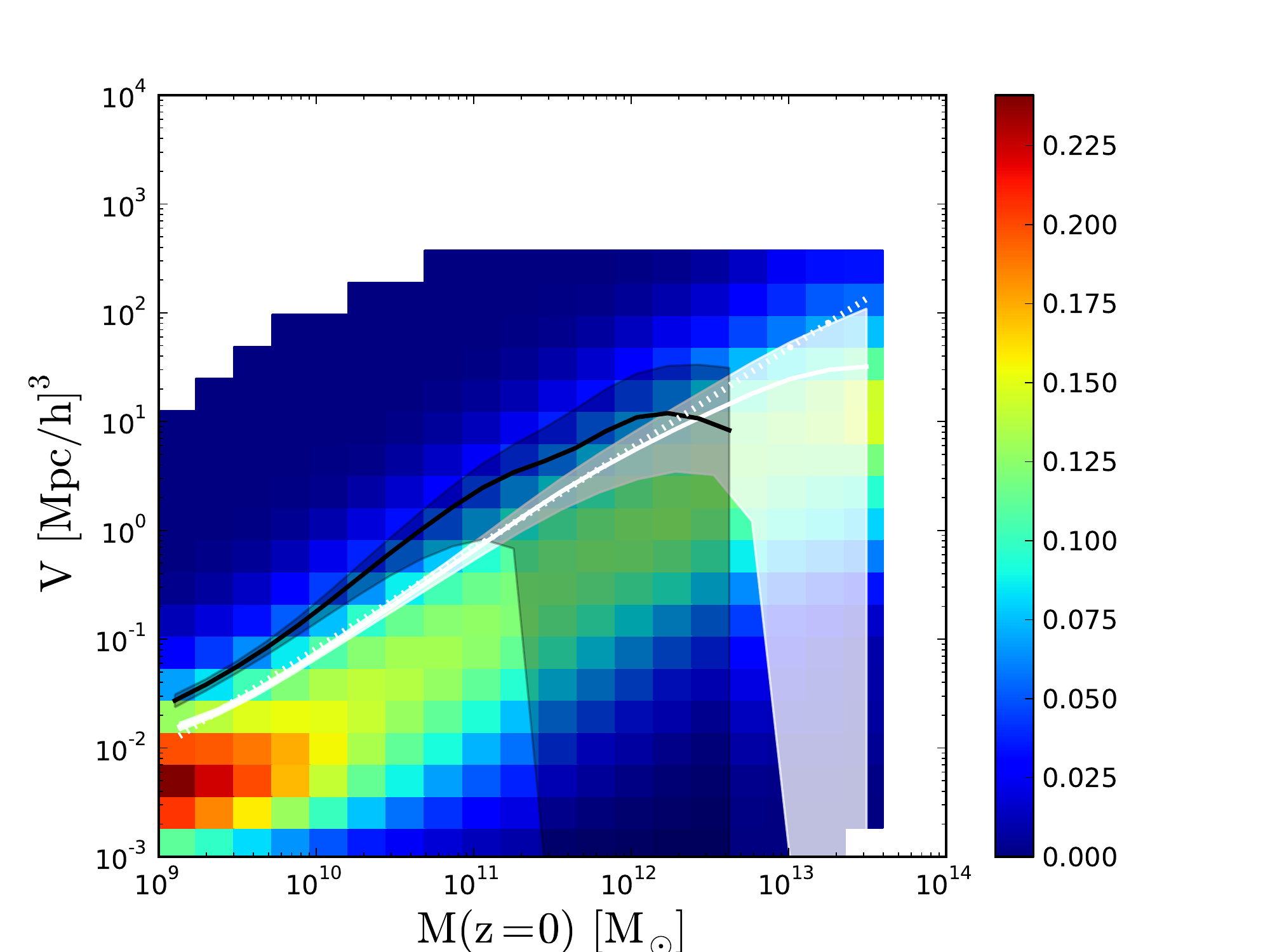} \\
      (c) S50 & (d) H50\\
\end{tabular}    
\caption{ In background, the distribution of the volume of initial HII region growth $t_\mathrm{life}$ for a halo of current $M_0$ mass, 
extrapolated from its value when the HII region experiences a major merger. 
The color code is in arbitrary unit with blue values indicating a faint probability while the red tones denote a high probability.
The white curve represents the evolution of the mean value of the whole distribution and the shaded area stand for the 3$\sigma$ uncertainty on this value.
The dotted white line represents the best fit of the mean value according to the fitting formula described in appendix \ref{fit2}.
In black the same average relation extrapolating the halo mass from the moment the HII region appears.}
    \label{volume_vs_mzero}
  \end{center}
 \end{figure*}

\begin{figure*}
   \begin{center} 
  \includegraphics[width=8cm,height=7cm]{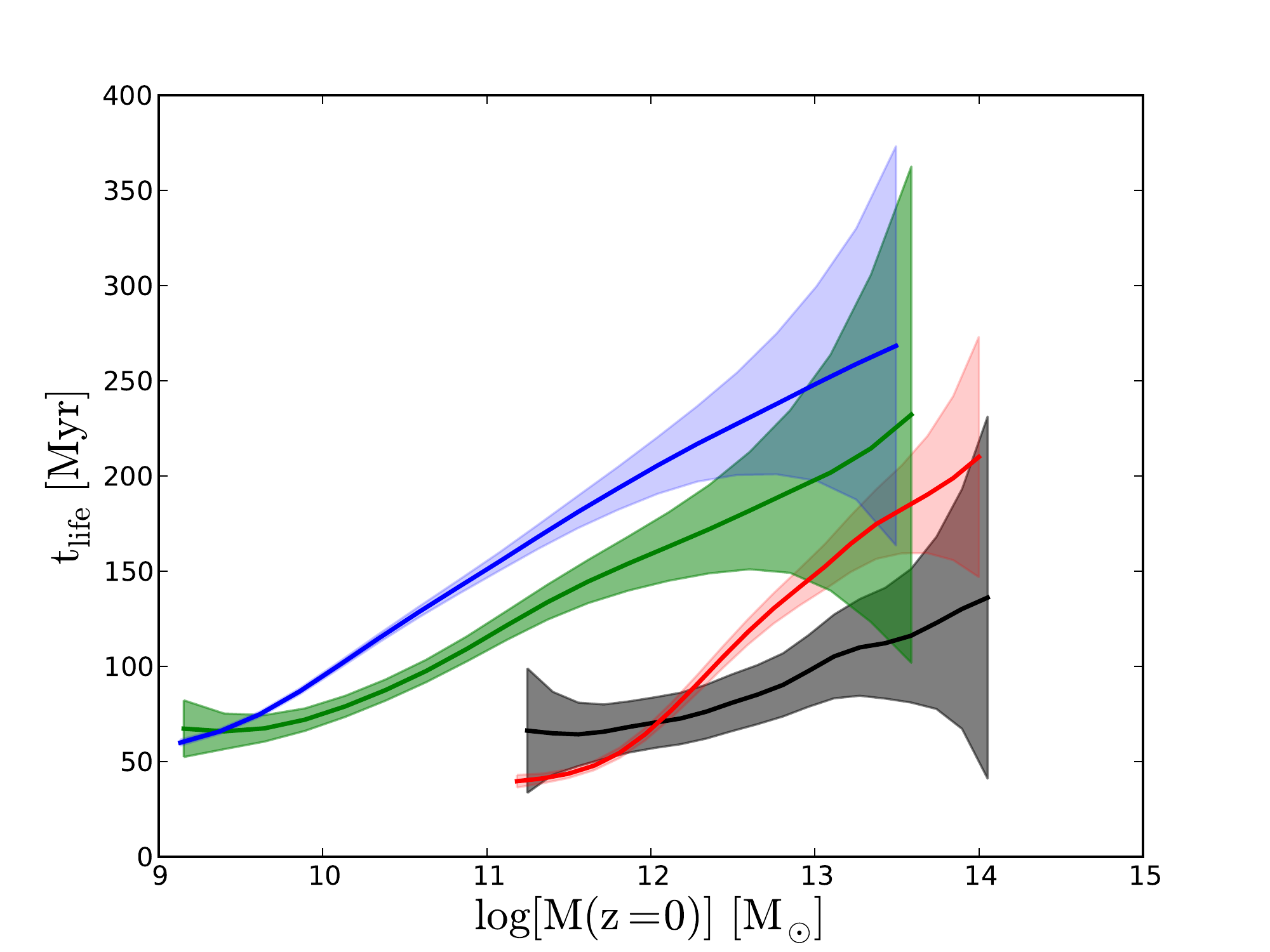} 
 \includegraphics[width=8cm,height=7cm]{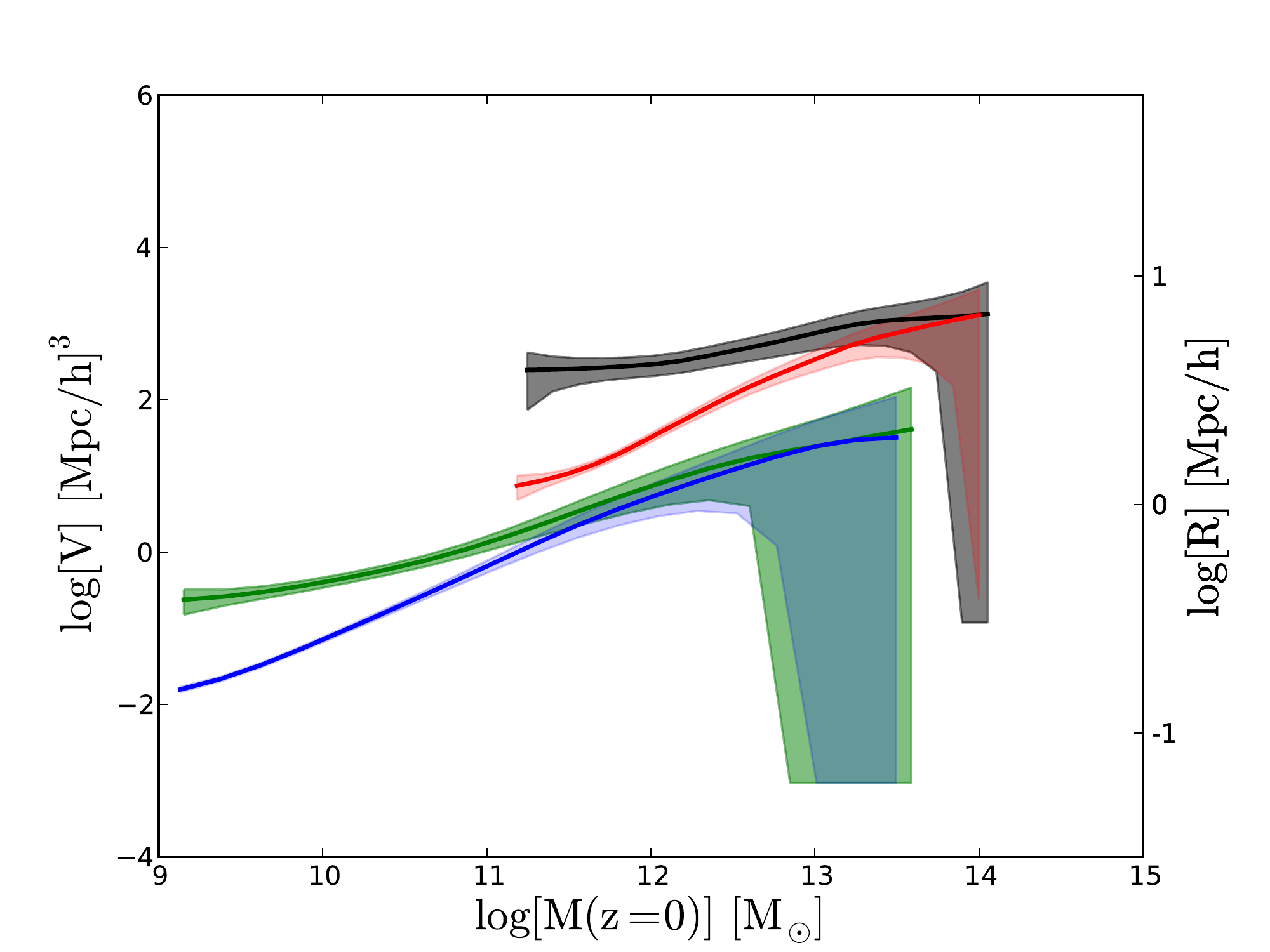} 
   \caption{Average duration (left) and maximal expansion (right) of the initial HII region growth as a function of the mass of the most massive halo extrapolated at z=0. 
Black and red curves stand respectively for the star and halo based model in the 200 Mpc/h simulation. Green and blue stand for the star and halo based model 
in the 50 Mpc/h simulation. Shaded areas stand for the $3\sigma$ error on the avergae value.}
    \label{allcurve}
  \end{center}
 \end{figure*}

We present the evolution $\mathrm{t_{life}(M_0)}$ of the lifetime of the regions before their major merger as a function of $\mathrm{M_0}$ in figure \ref{lifeteimevsmzero}.
The white curves show the evolution of the mean value per bin of $\mathrm{M_0}$ as well as the distribution in background. Shaded areas stand for the $3\sigma$ error on the mean value. 
Using the same conventions, figure \ref{volume_vs_mzero} shows the final volume of an isolated HII region as a function of today's mass, $\mathrm{V(M_0)}$ . For comparison, all the models 
are superimposed in figure \ref{allcurve} for both $\mathrm{t_{life}(M_0)}$ and $\mathrm{V(M_0)}$.

Comparing the source models for a fixed resolution (i.e. S50 Vs H50 or S200 Vs H200), the same trends can be observed. At large $\mathrm{M_0}$, the final volume of HII regions 
does not depend on the model. This convergence is achieved for $\mathrm{M_0>10^{13} M_\odot}$ in 200 Mpc/h simulations and $\mathrm{M_0>10^{11} M_\odot}$ in 50 Mpc/h ones. In terms of 
isolation duration ($t_\mathrm{life}$), the same objects will nevertheless differ : halo-based models tend to produce longer local reionization than in star-based models. 
Light objects follow quite an opposite trend: halo-based model tend to surround this objects with smaller HII regions than star-based ones and at the same time the differences 
in duration of local reionizations tend to be reduced. Let us recall that emissivities were tailored to have both models that produce the same global reionization history and 
therefore similar photon production histories. Let us also recall that halo models provide more sources than star-based models: as a consequence the number of photons cast 
per halo is smaller at a given resolution. For large mass objects, the final volume does not depend on the source model, suggesting that large HII regions are spatially 
distributed in the same manner in both descriptions. At the same time emissivities are smaller in H models leading to slower propagation of I-Fronts than in S calculations. 
For light objects, the situation is slightly different: H models present a large number of small and clustered HII regions, induced by small halos during the reionization 
that translate into small objects at z=0. Therefore the final volumes are smaller than in star-based models, compensate for the smaller emissivities per source and 
reduce the difference in terms of duration of local reionizations. This differences between H and S models were already present in the merger-tree and radii statistics 
in \cite{2012A&A...548A...9C} and the results obtained here is another view of the impact of source modeling on the geometry of HII regions during the reionization.

Decreasing the simulated volume leads typically to a decrease in HII regions volumes and an increase in the duration of HII regions expansion. Again, increasing the 
resolution increases the number of sources for both models and therefore decreases the photon production rate per source. The larger density of sources reduces naturally 
the volume available for an HII region and in order to have similar global reionization histories in 50 and 200 Mpc/h simulations, the expansion rate of HII regions 
must be reduced in 50 Mpc/h experiments. Quantitatively, the final volumes are one order of magnitude greater in 200 Mpc/h simulations with typical radii of 5.2 Mpc/h 
for the HII regions before they merge for $\mathrm{10^{13} M_\odot}$ haloes against 1.7 Mpc/h in 50 Mpc/h experiments.  For the least massive halos within the 50 Mpc/h simulation 
($\mathrm{M_0 \sim 1.5\times 10^9 M_\odot}$), their radii can be as small as 150 kpc/h in the H50 model  and 350 kpc/h in the S50 model. The difference is greater for light objects 
in the larger box with a typical radii of 1.3 Mpc/h in H200 for $\mathrm{M_0 \sim 2\times 10^{11} M_\odot}$ and 4Mpc/h in S200. In terms of duration, $\mathrm{t_{life}}$ increases from 
70 Myr to 250 Myr in 50 Mpc/h experiments and from 70 to $\sim 200$ Myrs in 200 Mpc/h but over a range of larger masses. Globally the two models are in better agreement in the small box, 
a result that could already be seen in the merger tree statistics of \cite{2012A&A...548A...9C}. It should be mentioned that looking at the dispersion in the distributions of values 
in backgrounds of figures \ref{lifeteimevsmzero} and \ref{volume_vs_mzero}, the scatter is still quite large in all situations, and the differences noted above lie within 
values that are allowed by this dispersion.

Whichever model or resolution is considered, the properties of the initial growth is strongly mass dependent. The most massive objects seen today grow from the earliest 
progenitors with an accretion rate then at the highest levels and a longer accretion history. Also the initial stages of the growth of their HII regions occurred within a 
mostly neutral Universe. Combined with the fact that they are strong emitters, it is thus expected that their HII regions would dominate for a long time before being merged into 
a larger UV background. In our models 300 Myr represents approximatively $1/3$ of the reionization epoch, i.e. a significant fraction,  corresponding to several dynamical 
times at this epoch or a few generation of massive stars. Taking $\mathrm{M_0=10^{12} M_\odot}$ as an example and depending on the box-size/model used to constrain these values, such a 
halo would have reionized a radius of $2.9\pm1.6$ Mpc/h after a period between 50 and 200 Myrs. It implies that for such a duration these objects experience inside-out reionizations with 
properties decoupled from a cosmic averaged behaviour, with for instance local anisotropies or inhomogeneities.  At the light end of the mass range explored here, 
the associated HII regions are merged into the UV background after a few tens of Myrs, resulting both from environmental effects where close larger regions dominate 
and from the fact that these objects appear at the latest stages of reionization where the ionized filling fraction is close to 1. These light objects experienced a 
reionization that is more in adequation with the usual picture of an object that is rapidly part of a uniform UV background. Overall, it is clear that depending on 
the mass (which is related to a variety of environment or a variety of internal history of source buildup) a large variety of local reionization histories are possible, 
quite different from the homogeneous rise of a global UV flux.

Finally, we show on the same figures, the relations obtained using a different procedure to compute the current mass $\mathrm{M_0}$ : instead of using the mass of the most massive 
halos as the HII region merge in the halo growth model, we use the same mass but when the HII region \textit{appeared}. The relations are shown in black in figs 
\ref{lifeteimevsmzero} and \ref{volume_vs_mzero}. It can be seen that it does not really make a significant difference, with trends and quantitative values consistent 
with the other choice of initial mass. The strongest impact is on massive halos which does not come as a surprise: their apparition occur at the earliest time, therefore the initial 
uncertainty on their mass (which is close to 10 particles) is propagated and amplified over a longer time.  Overall it acts as an indirect proof that our arbitrary choice of $\mathrm{M_f}$ 
does not have a strong influence on the extrapolation.

\subsection{Apparition time and redshift of merger with the UV background as a function of $\mathrm{M(z=0)=M_0}$}
\label{ltimebmajor2}

\begin{figure}[tb]
   \begin{center}
      \includegraphics[width=9cm,height=7cm]{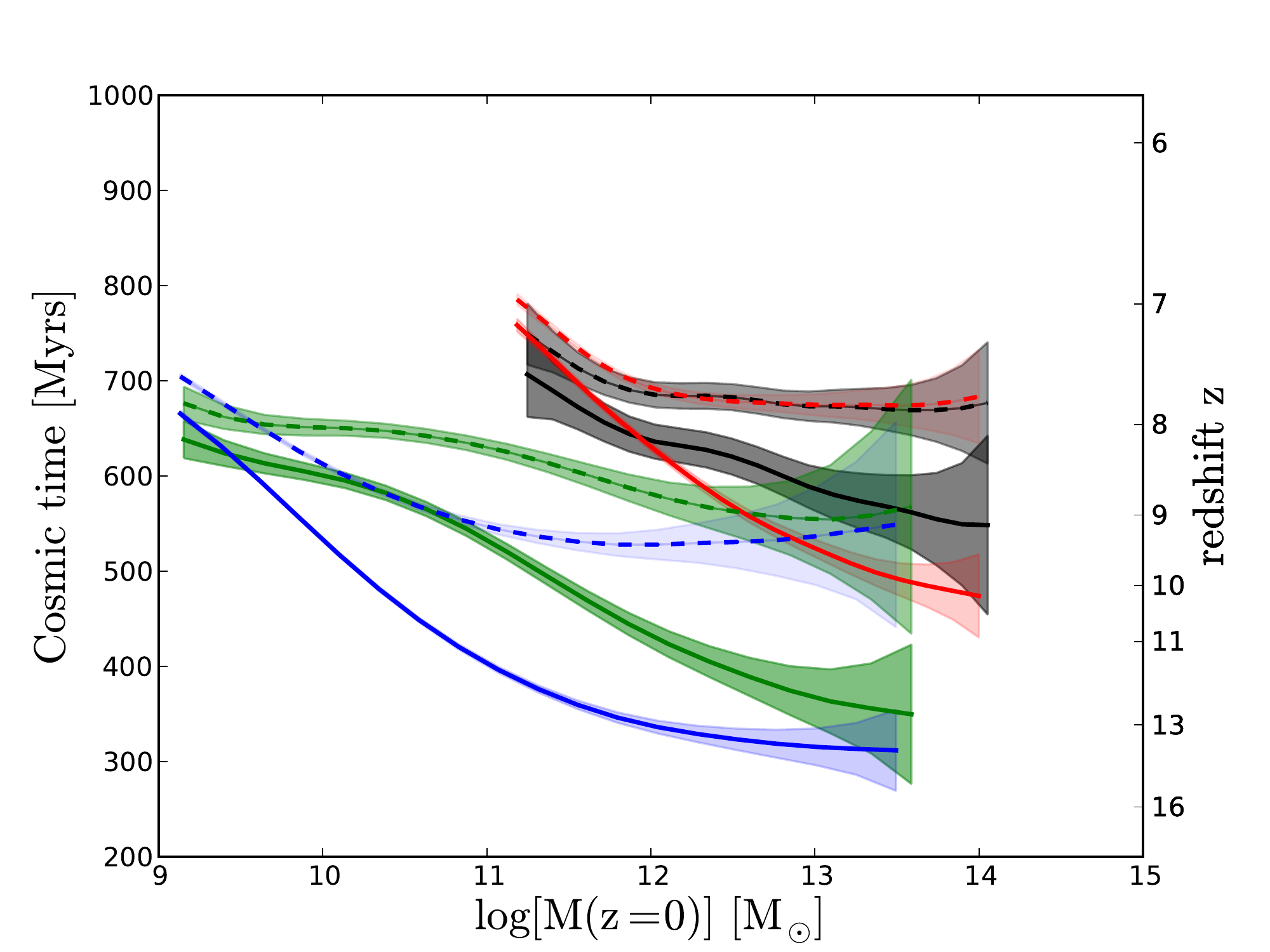} \\
  \caption{Mean cosmic apparition time (solid lines) and mean redshift of merger with the UV background (dashed lines) as a function of the mass of the most massive halo extrapolated at z=0. 
Black and red curves stand respectively for the star and halo based model in the 200 Mpc/h simulation. Green and blue stand for the star and halo based model 
in the 50 Mpc/h simulation. Shaded areas stand for the $3\sigma$ error on the average value.}
    \label{tapp_vs_m0}
\end{center}
\end{figure}

In figure \ref{tapp_vs_m0}, we show the evolution of the mean cosmic apparition time of the HII regions as a function of the mass of 
their progenitors extrapolated at z=0 for our four models.
It is first reassuring to note that, whatever the model considered, the most massive haloes today are on average associated with the oldest HII regions, with the smallest apparition times.
The different HII regions can appear as soon as 300 Myrs after the Big-Bang in the 50 Mpc/h box for the most massive halos observed at z=0.
The last appearing regions corresponding to light z=0 halos can emerge until 650 Myrs in the same box.
The first regions appear later on in the large box of 200 Mpc/h at $\sim 500$ Myrs but last regions emerge until $\sim 750$ Myrs.
These differences reflect the differences of resolutions with sources appearing earlier in the 50 Mpc/h boxes compared to the 200 Mpc/h experiments 
where sources appear later but can reach larger values of apparition cosmic time in order  to complete the reionisation.  

On the same figure, we also show  the evolution of the mean cosmic time (redshift) of merger with the UV background for the halos as a function of their mass at z=0.
Surprinsingly, we observe a sort of `plateau' in all the models for the reionization redshift over a certain range of $M_0$ halos.
Indeed, an almost constant redshift of $z \sim 8-9$ is observed respectively in the 200 and 50 Mpc/h boxes for halos with $M_0>1\times 10^{12} M_\odot$.
These redshifts corresponds to the beginning of the intense merger period of the HII regions and to the emergence of a main HII region in size as noted in \cite{2012A&A...548A...9C}. 
It thus seems that halos with $M_0>1\times 10^{12} M_\odot$ initiated their build-up before that period, that their associated HII regions dominated the merging process and that, on average, they finished their isolated reionisation history at the intense overlap epoch whatever their apparition time. 
  
Halos with  $M_0<1\times 10^{12} M_\odot$ initiated on average their build-up during or after this merger period and thus in an already significantly ionised environment.
We observe that the lighter the halos, the later it reaches the reionisation with a difference between its lifetime and its apparition time which is decreasing.
This is expected as soon as the lighter halos are those appeared the later with an enhanced proximity effect with other HII regions.

\section{Discussion}
\label{discussion}

\subsection{Convergence issues}
\label{valid_results}

The previous results put into light one important fact which is the difficulty to reach the convergence between the two models of ionizing sources at both resolution.
Despite the certain degree of similaritiy reported in \cite{2012A&A...548A...9C} (paper I) between the two models, only the same global behavior can be observed.
All the ionizing source models and resolution studies agree on the fact that the most massive z=0 haloes have HII regions expanding during a much longer time and thus 
filling in a much larger volume than light objects before being incorporated into the UV background.
However the quantitative results are far to be close and extremely model and resolution dependant.
Nevertheless, we want here to show that some halo mass ranges exist, depending on the resolution, where our results are converged for both models of ionizig sources.
The main problem with our quantitative results shown in the next is the advantages and drawbacks concerning the two resolutions cases studied.
Indeed, the 200 Mpc/h box has the advantage of resolving the rare density peak, thus accounting for cosmic variance effect and allowing to track the largest HII regions.
Differently, the 50 Mpc/h has a better resolution than its 200 Mpc/h analog and can thus resolve smaller HII region and account for the inhomogeneity of the reionisation process.
We have to keep in my mind this aspects when interpreting our results in the next.

In paper I, we noted that both ionizing source prescription lead 
to similar evolution for the radius distribution of HII regions with redshift and for their merger history. 
It thus appears that star based model could be able to reproduce at a certain level a similar morphology 
of the reionization process than the one provided by halo sources.
With the present study it seems that this affirmation seems to loose a degree of validity when extrapolating the results from high-z to z=0.

More precisely, we found in paper I (figure 11) a cutoff radius (0.4 Mpc/h (resp. 4 Mpc/h) in the 50Mpc/h model (resp. 200 Mpc/h)) 
in the evolution with redshift of the radius distribution of HII regions. 
Above these radius cutoff, the radius distribution of the star prescription becomes the analog of the one providied by the halo sources.
In the $\mathrm{V(M_0)}$ evolution of figure \ref{allcurve}, we therefore have a better robustness of the results in the range of volume above these cutoffs at both resolution.
This is the volume range where we argue that our results converge for both ionizing source prescriptions.
At both resolution these `converged' volume ranges can translate into z=0 halo mass range.
Indeed, we can consider only the mass range where both $\mathrm{V(M_0)}$ curves corresponding to the two ionizing source model are above the cutoff radius.  
Therefore the mass ranges for which we are the most confident regarding our results are 
for haloes with mass $\mathrm{M\ge 1\times10^{13} M_\odot}$ (resp. $\mathrm{M\ge 1\times10^{11} M_\odot}$) in the 200 (resp. 50 Mpc/h box).

In the present paper, when we consider the $\mathrm{V(M_0)}$ evolution for the mass range of confidence, 
we can note that the convergence in the reionized volume by the haloes is still present between both ionizing source models: 
the evolution of the HII regions size becomes almost superimposed in the two ionizing source model at both resolutions.
In other words, the convergence seen at high z for the HII regions size is translated at z=0.
However it seems that differences appear in the $\mathrm{t_{life}(M_0)}$ between the two models at both resolutions.
In the two resolution cases, a difference of $\sim$ 50 Myr in the isolated reionisation history duration is observed almost all along the halo mass ranges of confidence.
For example, a halo with a mass of $\mathrm{\sim 1\times10^{13} M_\odot}$ has a $\mathrm{t_{life}\sim 100-150}$ Myr 
respectively for the star and halo model in the 200 Mpc/h experiments and $\mathrm{t_{life}\sim 200-250}$ Myr in the 50 Mpc/h simulations.      
This gap reflects again the difference between the emissivity of the two models with a high number of haloes casting a smaller number of ionizing photons
than the stronger star emitters needed to counterbalance their smaller number. 
In other words we observe a memory effect of the differences generated by the emissivity models that are still present in the z=0 results.

\subsection{Predictions for the Local Group}
\label{LGroup}

In the 200 Mpc/h simulations, the halo mass range where we get a good level of convergence between the two models begins with 
too high masses (above $\mathrm{\sim 10^{13} M_\odot}$) to make strong predictions 
about the past reionization history of peculiar galaxies like the Milky Way with a halo mass of $\mathrm{\sim 10^{12} M_\odot}$ 
(see \citealt{2005MNRAS.364..433B}).  
However in the 50 Mpc/h experiment, we get a reasonable degree of convergence between the simulation for halos with mass $\mathrm{M \ge \times10^{11} M_\odot}$.
At this resolution, $\mathrm{10^{12} M_\odot}$ haloes have reionization duration of $\sim 150 \pm 15$ Myrs or $\sim 200 \pm 15$ Myrs 
respectively in the star and halo prescriptions and a typical radius of $\sim$ 1.1 Mpc/h for the HII region.
Recently, in \cite{2014ApJ...785..134L} a study was conducted in order to assess the reionization history of Milky Way-type halos. 
They found median reionization times of $\sim$ 115 Myr for halo mass of $\mathrm{1\times10^{12} M_\odot}$ which is a little lower than our averaged estimate.
Nevertheless, they used median values which could be close to our results because our averaged value is greater than the median estimate we could get.  
On the other hand, in \cite{2013ApJ...777...51O}, radiative transfer simulations in the CLUES constrained realizations (see \citealt{2010MNRAS.401.1889L}) of the Local Group
produce isolated reionizations of the Milky Way ($\mathrm{M \sim 3\times10^{11} M_\odot}$ ) for 130 Myrs in their photon-rich H43 model (the closest to our emissivity model) 
with a maximal extension of $\sim$ 1 Mpc/h. 
In our case, for the same halo mass, we find $\sim 135 \pm 10$ and $\sim 175 \pm 10$ Myrs respectively for the star and halo models with a typical radius of $\sim$ 0.7 Mpc/h
which is close to their results.
It thus seem that our statistical sample of reionisation histories could be representative of the particular case of the MW reionization history.

\begin{figure*}
   \begin{center} 
  \includegraphics[width=8cm,height=17cm]{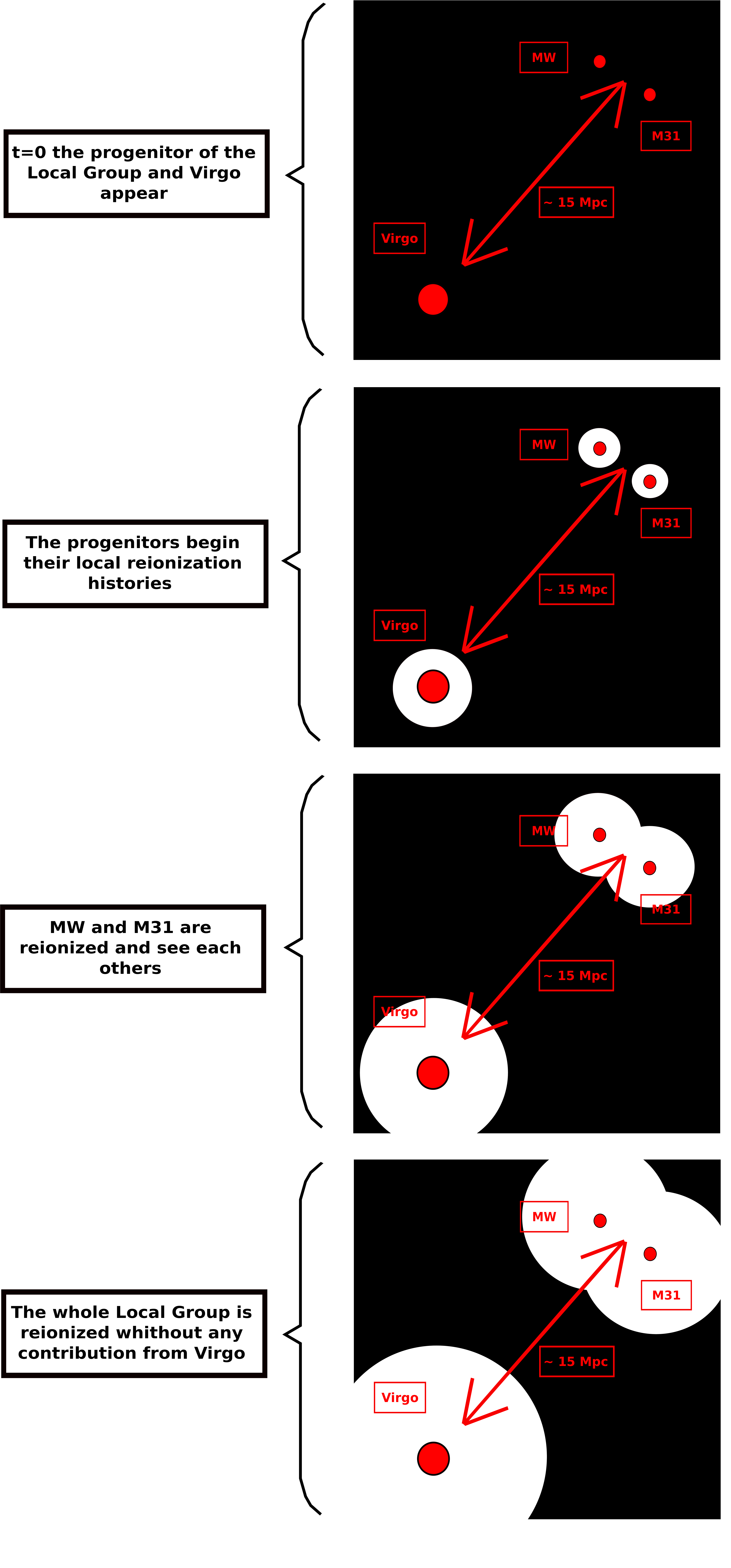} 
   \caption{Ilustration of the reionization of a Local Group-type object as suggested by our results.}
    \label{mw_m31_local_group_virgo}
  \end{center}
 \end{figure*}

In figure \ref{mw_m31_local_group_virgo}, we represent an illustration of the reionization process of a Local Group-type objects as suggested by our results.
Thinking of a halo of  $\sim \, \mathrm{3\times10^{12} M_\odot}$ (see \citealt{2002ApJ...573..597K}) with a mass comparable with the whole Local Group (composed of MW and M31), 
our results indicate a possible HII region extension with a radius of $\sim 1.5$ Mpc/h (still in the 50 Mpc/h box) before encountering another front. 
In other words a volume just large enough to encompass the whole Local Group. 
\textit{Statistically} a halo with a mass comparable to those of the Local Group could therefore have been able to 
reionize by itself and not be swept by the radiation of a Virgo-like close cluster.

However, the Local Group could also been reionized by dwarf galaxies instead by their central lighthouse according to large scale simulation of cosmic reionization 
(see \citealt{2007MNRAS.376..534I}, \citealt{2011MNRAS.414..847S}, \citealt{2012MNRAS.423..862K} and \citealt{2013MNRAS.428L...1M}).
Such studies suggest that haloes that would form the dwarf galaxies with total masses $M \le 10^9 \mathrm{M_\odot}$  could provide a significant fraction of the ionising photon 
budget during cosmic reionisation (\citealt{2014arXiv1403.6123W}).
Cosmic reionisation could indeed have been primarily driven by proto-galaxies in haloes with masses between $\mathrm{10^{7} M_\odot}$ and $\mathrm{10^{8} M_\odot}$, 
which have very high escape fractions (\citealt{2013MNRAS.429L..94P}).
Moreover, \cite{2014MNRAS.437L..26S} suggests that dwarf galaxies with M $\le \mathrm{10^{9} M_\odot}$  could provide a significant fraction of
the ionising photon budget during cosmic reionisation and that
M $\le \mathrm{10^{9} M_\odot}$ dwarf galaxies could have dominated the reionization of galaxies like our own Milky Way providing $>$ 80\% of the required photons budget.
Unfortunately, in our case, we do not have the resolution to track such dwarf galaxies. 
Nevertheless, recent studies of \cite{2013ApJ...777...51O} would suggest alternately
that the reionization of the Local Group could have been driven in an inside-out way under the influence of the central main halos.
In fine, the answer to the question of an internal versus external reionization history of MW-type galaxies could be tested with higher 
resolution large scale simulations of the Local Group.

Finally, as argued before, we want to stress that the 50 Mpc/h simulations has the default of finite volume effects and cannot 
therefore include rare density peaks or large voids that may be expected.
Moreover it has not a volume large enough to track the evolution of large HII regions during the reionization.
To strengthen the results drawn from our 50 Mpc/h experiment, the ideal case would be to
have the 200 Mpc/h simulation with the same resolution than the 50 Mpc/h one to be sure that we are insensitive to cosmic variance.

\section{Summary and prospects}
\label{prospects}

We investigate simple properties of the initial stage of the reionization process around progenitors of galaxies, such as the extent of the initial HII 
region before its absorption by the UV background and the duration of its propagation.
We have investigated the multiple reionization histories in a set of four simulations where we have varied the ionizing source prescriptions and the 
resolution of the simulations. By using a merger tree of HII regions (here defined as regions with a ionising fraction x $\ge 0.5$), 
we made a catalog of the HII regions properties  for all ionized regions appeared in the experiments.
We then looked at the time when the HII regions undergo a major merger event, after then we considered that the region is apart of the global UV background.

By looking at the lifetime of the region and their volume at this moment we draw typical local reionization histories as a function of the cosmic time of 
appearance of the HII regions. We also investigate the link between these histories and the dark matter halo masses inside the regions. By using a motivated 
functional form for the average mass accretion history of the dark matter halos, we then extrapolate the mass inside the region at $\mathrm{z=0}$ in order to 
make predictions about the past reionization histories of galaxies seen today.

Our results can be summarized as follows:

\renewcommand{\labelitemi}{\textbullet}
\begin{itemize}
 \item{We found that the later an HII region appears during the reionization period, the smaller will be their related lifetime and volume before they see the global UV background. 
This is a normal consequence of the overlap of ionized patches that reduce the neutral volume available for I-front propagation. However, quantitatively the duration and the 
extent of the initial growth of an HII region is strongly dependent on the mass of the inner halo. During this initial stage, that can be as long as 350-400 Myrs (i.e $\sim 50\%$ of the reionization),  
this inner halo and the galaxies it hosts is decoupled from the external UV background.}
\item{We extrapolate the mass of dark matter halos at high z to z=0 using a halo growth model to predict the extent and the duration of this initial HII regions around current galaxies. 
The quantitative prediction differs depending on the box size or the source model: while enforcing similar global reionization histories, small simulated volume promote proximity effects 
between HII regions and halo-based source models predict smaller regions and slower I-front expansion.} 
\item{By looking at the mean reionisation redshift of the galaxies as a function of $M_0$, we find that, on average, halos with $M_0>1\times 10^{12} M_\odot$ 
appeared during the pre-overlap period
and quit the isolated reionisation regime at the overlap whatever their apparition time in the cosmic history.}
\item{Applying this extrapolation to the Local Group leads to a typical extent of 1.1 Mpc/h for the initial HII region around a typical Milky Way that established 
itself in $\sim 150-200\pm20$ Myrs. 
This is comparable with recent constrained simulation of the Local Group done by \cite{2013ApJ...777...51O} telling us that our statistical study could be representative of
particular galaxy ionization histories in this mass range. 
Considering the whole Local Group, our result suggests that statistically it should not have been influenced by an external front coming from a Virgo-like cluster.}
\end{itemize}

From the results of this study, we plan to repeat this technique on enhanced simulations of reionization, the ideal situation being an experiment 
where at least the dark matter field evolution would be run until $\mathrm{z=0}$.
This could make us free from the impact of any semi-analytical framework for the extrapolation of the mass 
at $\mathrm{z=0}$ and we therefore would reason only with quantities directly generated during the simulation.
Using the same methodology, we aim at expanding this kind of measures to the statistics of the inner properties 
of these local reionizations such as i-front propagation profiles or anisotropies. 
Ultimately, they would provide a detailed framework of the initial stage of reionizations around galaxies that 
could be included in semi-analytical modeling or hydrodynamical simulations without radiative transfer. 

\section*{Acknowledgments}

The authors would like to thank Herv\'e Wozniak, Beno\^it Semelin and 
Martin Haehnelt for comments and discussions. 
The simulations were run on the Curie Supercomputer (CCRT-CEA) using PRACE Preparatory Access time.
This study was performed in the context of the EMMA (ANR-12-JS05-0001) \& 
LIDAU (ANR-09-BLAN-0030) projects, funded by the Agence Nationale de la Recherche (ANR).

%
\bibliographystyle{aa}
\bibliography{biblio}

\appendix

\section{Fitting models}
\label{fit1}

To approximate the cosmic time dependence of the mean lifetime of the HII regions, we use two parameters exponential form:

\begin{center}
\begin{equation}
 \mathrm{t_{life}=ae^{bx}}
\end{equation}
\end{center}

In table \ref{tab3} we summarize the best fits parameters for the mean curve of the complete distribution for each simulation. 
We represent the best fits of the curves in figure \ref{lifeteimevstime} with the dotted white line.

To approximate the cosmic time dependence of the mean volume of the HII regions at the moment of the major merger, we use two parameters exponential form:

\begin{center}
\begin{equation}
 \mathrm{V=ae^{bx}}
\end{equation}
\end{center}

We report in table \ref{tab3} the best fits coefficients found in every models and give also the mean dispersion along 
the curve and the best fits of the curves are shown in figure \ref{volume_vs_time} with the dotted white line.
Such fits could be useful in order to assign typical volume for the HII regions as a function of the cosmic time.

\renewcommand{\arraystretch}{1.5}
\setlength{\tabcolsep}{0.5cm}
\begin{table*}[tb]
\begin{center}
\begin{tabular}{|c|c|c|c|c|}
\hline
  Model name & $<t_{life}>$ & $\sigma_{t_{life}}$ & $<Volume>$ & $\sigma_{volume}$  \\
  \hline
  S200 & a=$1398$  b=$-4.76\times 10^{-3} $ & $6.15 $   &  a=$8.31\times 10^{4}  $  b=$-8.04\times 10^{-3} $ & $363.5 $\\

  H200 & a=$1367$  b=$-4.66\times 10^{-3}$ & $3.11 $ &  a=$1.38\times 10^{5}  $  b=$-1.27\times 10^{-2}$ & $84.6 $   \\

  S50  & a=$517.1$   b=$-3.09\times 10^{-3} $ & $3.16 $ &   a=$26.0 $  b=$-5.81\times 10^{-3}$ & $0.63 $   \\

  H50  & a=$330.8  $  b=$-2.36\times 10^{-3}$ & $1.04 $  &  a=$7.02  $  b=$-8.20\times 10^{-3}$ & $0.04 $  \\
  \hline
\end{tabular}
\caption{Linear fits for the cosmic time dependence of the mean lifetime $\mathrm{<t_{life}>}$ of a region before its first major merger and its mean volume $\mathrm{<Volume>}$ at this moment.}
\label{tab3}
\end{center}
\end{table*}

\section{Fitting models at z=0}
\label{fit2}

To approximate the $\mathrm{M_0}$ dependence of the mean lifetime of the HII regions, we use two parameters form:

\begin{equation}
 \mathrm{t_{life} = alog10(M_0)+b}
\label{tlife_vs_m0_form}
\end{equation}

In table \ref{tabtlife_volume_vs_M0} we summarize the best fits parameters for the mean curve of the complete distribution for each simulation. 
We represent the best fits of the curves in figure \ref{lifeteimevsmzero} with the dotted white line.

To approximate the $\mathrm{M_0}$ dependence of the mean volume of the HII regions at the moment of the major merger, we use two parameters form:

\begin{equation}
 \mathrm{V = a M_0^b}
\label{Volume_vs_m0_form}
\end{equation}

We report in table \ref{tabtlife_volume_vs_M0} the best fits coefficients found in every models and give also the mean dispersion along 
the curve and the best fits of the curves are shown in figure \ref{volume_vs_mzero} with the dotted white line.
Such fits could be useful in order to assign typical volume for the HII regions as a function of the cosmic time.

\renewcommand{\arraystretch}{1.5}
\setlength{\tabcolsep}{0.5cm}
\begin{table*}[tb]
\begin{center}
\begin{tabular}{|c|c|c|c|c|}
\hline
  Model name & $<t_{life}>$ & $\sigma_{t_{life}}$ & $<Volume>$ & $\sigma_{volume}$  \\
  \hline
  S200 & a=$26.9$  b=$-248.9$ & $9.1$   &  a=$3.59\times 10^{-2}$  b=$0.33$ & $145.9$\\

  H200 & a=$67.1$  b=$-728.0$ & $4.6$ &  a=$5.15\times 10^{-11}   $  b=$0.99$ & $75.9$ \\

  S50  & a=$39.1$   b=$-310.0$ & $8.8$ &   a=$1.87\times 10^{-8}$  b=$0.72$ & $4.45$   \\

  H50  & a=$50.25$  b=$-403.6$ & $6.6$  &  a=$5.01\times 10^{-11} $  b=$0.92$ & $3.34$  \\
  \hline
\end{tabular}
\caption{Linear fits for the $M_0$ dependence of the mean lifetime $\mathrm{<t_{life}>}$ of a region before its first major merger and its mean volume $\mathrm{<Volume>}$ at this moment.}
\label{tabtlife_volume_vs_M0}
\end{center}
\end{table*}

\end{document}